\documentclass[
  aps,
  reprint,
  superscriptaddress,
  amsmath,
  amssymb,
  floatfix
]{revtex4-2}

\usepackage{bm}

\usepackage{graphicx}
\usepackage{dcolumn}
\usepackage{booktabs}
\usepackage{makecell}

\usepackage{algorithmic}

\newcounter{algorithm}
\renewcommand{\thealgorithm}{\arabic{algorithm}}

\newenvironment{revtexalgorithm}[2][]{%
  \refstepcounter{algorithm}%
  \begin{table}[#1]
  \footnotesize
  \setlength{\tabcolsep}{0pt}
  \begin{ruledtabular}
  \begin{tabular}{@{}l@{}}
  \begin{minipage}{0.96\columnwidth}
  \textbf{Algorithm~\thealgorithm.}~#2\par
  \vspace{2pt}
  \hrule
  \vspace{3pt}
}{%
  \vspace{2pt}
  \end{minipage}\\
  \end{tabular}
  \end{ruledtabular}
  \end{table}
}

\usepackage[colorlinks=true,
            linkcolor=blue,
            citecolor=blue,
            urlcolor=blue]{hyperref}

\begin{document}

\title{
Sequential Spatiotemporal Magnetic-Field Reconstruction via
Quantum Hamiltonian Learning with NV-Center Spin-1 Hamiltonians
}

\author{Hiroshi Yamauchi}
\email{hiroshi.yamauchi@g.softbank.co.jp}
\affiliation{SoftBank Corp.}

\author{Sophie Colleen Stearn}
\email{sophie.stearn@quantum-brilliance.com}
\affiliation{Quantum Brilliance}

\author{Samuel Tovey}
\email{samuel.tovey@quantum-brilliance.com}
\affiliation{Quantum Brilliance}

\begin{abstract}
We propose a quantum-Hamiltonian-learning-based sequential reconstruction
framework for dynamic two-dimensional magnetic-field maps using a local
likelihood model derived from a nitrogen-vacancy-center spin-1 Hamiltonian.
Local measurements are generated through nitrogen-vacancy spin dynamics
governed by local magnetic-field values and a shared dipolar coupling
parameter, rather than by direct observation of the latent field.
Sequential Bayesian updates over overlapping scan windows are combined with
temporal posterior propagation to reconstruct the evolving field.

Numerical proof-of-concept experiments on controlled synthetic maze-like
magnetic-field sequences show that the proposed method reconstructs the
dominant spatial structure of the tested field class, achieving a final-frame
RMSE of \(7.037\times10^{-7}\,\mathrm{T}\). Adaptive diagnostics show
decreasing expected information gain and stable local convergence, while
Fisher-information and leakage diagnostics reveal a sensitivity--leakage
tradeoff under long-interrogation controls.

Combined horizontal and vertical scans yield better reconstruction than
single-direction acquisition in the tested setting. In contrast, the shared
coupling parameter \(J\) is only partially identifiable: its posterior
becomes narrow but remains frame-dependent and biased. At the final
checkpoint, \(J_{\rm std}=87.0\,\mathrm{Hz}\), close to a finite-time
product-state reference benchmark of \(73.3\,\mathrm{Hz}\), while remaining
\(3.35\times\) above a gain-extrapolated ideal-state benchmark. The
posterior mean remains biased by \(326.9\,\mathrm{Hz}\), indicating that
posterior concentration alone does not imply unbiased coupling recovery.
These results demonstrate feasibility for the tested structured field class
and identify coupling estimation as the main identifiability bottleneck.
\end{abstract}

\keywords{
Quantum Hamiltonian Learning;
NV centers;
quantum sensing;
quantum metrology;
Fisher information;
Bayesian inference;
sequential estimation;
spatiotemporal reconstruction
}

\maketitle

\section{Introduction}

Reconstructing spatially distributed magnetic-field maps from incomplete
and noisy observations is a central problem in sensing, inverse modeling,
and data-driven scientific inference
\cite{KaipioSomersalo2005InverseProblems,Tarantola2005InverseProblemTheory,Stuart2010BayesianInverseProblems}.
In many practical situations, the target field cannot be measured directly
over the entire domain at once. Instead, observations are acquired locally,
sequentially, and under limited measurement resources, a setting naturally
related to sequential Bayesian filtering and particle-based state
estimation
\cite{Gordon1993BootstrapFilter,Arulampalam2002ParticleFilters,Doucet2001SMC}.
Moreover, the field of interest may evolve over time, so that the
estimation problem is inherently spatiotemporal rather than purely spatial.
In such settings, a reconstruction algorithm must recover local structure
from partial observations while propagating information across successive
measurements and maintaining consistency over multiple time frames. This
motivates sequential spatiotemporal field reconstruction, in which local
observations are assimilated progressively to recover a global
magnetic-field map that changes over time.

A natural physical platform for local field-sensitive measurements is
provided by nitrogen-vacancy (NV) centers in diamond
\cite{Doherty2013NVReview,Rondin2014NVBasedMagnetometry,Barry2020SensitivityOptimization}.
Hereafter, we use NV to denote nitrogen-vacancy centers. NV centers support
coherent spin dynamics that depend sensitively on the local magnetic
environment and therefore provide a physically grounded route to
likelihood-based field inference
\cite{Taylor2008DiamondMagnetometer,Maze2008NanoscaleMagneticSensing,Balasubramanian2008NanoscaleImaging}.
In distributed or scanning sensing scenarios, local measurements may be
acquired sequentially over a spatial domain while a hidden magnetic-field
pattern is reconstructed from the resulting measurement statistics.
Although the present work does not claim hardware-level deployment, this
setting provides a concrete motivation for studying reconstruction
frameworks that combine NV-based local quantum measurement models with
sequential global inference.

Classical approaches to dynamic field estimation are typically formulated
in terms of direct observation models, state-space systems, filtering
methods, or variational inverse problems
\cite{KaipioSomersalo2005InverseProblems,Tarantola2005InverseProblemTheory,Stuart2010BayesianInverseProblems,Doucet2001SMC}.
These methods have achieved broad success in applications ranging from
environmental monitoring and tomography to sequential image reconstruction
and sensor fusion. Nevertheless, they often assume that the measurement
process can be modeled directly in terms of the target variable, or through
a comparatively simple observation operator. In contrast, there are
settings in which the relation between the latent field and the observed
data is mediated by structured physical dynamics, producing nonlinear and
parameter-dependent measurement statistics. In such cases, reconstruction
is naturally viewed as a sequential inference problem governed by an
underlying dynamical model.

Quantum Hamiltonian Learning (QHL) provides a principled likelihood-based
perspective for this type of inference
\cite{Wiebe2014,Granade2012RobustOnlineHamiltonianLearning,Wang2017ExperimentalQHL}.
In QHL, unknown parameters are estimated by comparing observed measurement
statistics with likelihoods induced by candidate Hamiltonian models. Rather
than treating measurement outcomes as direct samples of the unknown
quantity, QHL interprets them as indirect evidence generated through
quantum evolution and measurement. This viewpoint is particularly suitable
when the latent parameter of interest affects system dynamics in a
structured way. By embedding unknown quantities into a Hamiltonian and
evaluating measurement likelihoods under candidate parameter values, one
obtains a Bayesian inference pipeline naturally suited to nonlinear
observation physics
\cite{Wiebe2014,Granade2012RobustOnlineHamiltonianLearning,Granade2016}.

While QHL has primarily been studied in the context of quantum system
identification, calibration, and parameter estimation
\cite{Wiebe2014,Granade2012RobustOnlineHamiltonianLearning,Wang2017ExperimentalQHL,KatoYamamoto2014SpinNetwork},
its use as a modeling principle for spatially distributed inference remains
less explored. Extending the setting from a small number of Hamiltonian
parameters to a time-varying spatial field introduces three main
challenges. First, the unknown quantity becomes high-dimensional and
distributed over space. Second, observations are available only through
local windows. Third, information must be propagated over time rather than
discarded after each local update. These considerations call for a
framework that combines Hamiltonian-based local likelihood modeling with
sequential Bayesian field estimation.

In this work, we develop such a framework and formulate spatiotemporal
magnetic-field-map reconstruction as a sequential inference problem based
on an NV-center Hamiltonian observation model. The target is a field map
defined over a spatial grid and indexed by discrete time frames. At each
step, the algorithm accesses the field only through local observation
windows. Each local window is associated with a parameterized quantum
measurement model in which local magnetic-field values determine the
underlying NV Hamiltonian and thereby modulate the resulting measurement
statistics. For each candidate field hypothesis, the model provides a
likelihood through Hamiltonian-induced dynamics and measurement
probabilities. These local likelihoods are then used in a sequential update
procedure that progressively refines the field estimate over space and
time.

The central idea is to bridge local inference and global reconstruction. A
local window contains only partial information about the full field, but
sequentially accumulated local observations can reveal larger spatial
structure when they are integrated consistently. Accordingly, the proposed
method does not attempt to infer the entire field from a single global
measurement model. Instead, it decomposes reconstruction into repeated
local likelihood evaluations and update steps, followed by aggregation into
a global field map. Repeating this process across multiple frames also
captures temporal evolution, allowing the reconstructed field to improve as
additional information becomes available.

A key feature of the present formulation is that the field is inferred
through measurement statistics induced by a physically motivated full
spin-1 NV Hamiltonian rather than by direct pixelwise observation.
Reconstruction therefore becomes a problem of identifying latent field
values from the statistical signature of NV spin dynamics. Because the
method is sequential, it also becomes possible to analyze not only the
final reconstructed field but also the intermediate stages through which
the estimate evolves. This is useful for understanding how local evidence
is incorporated, how spatial structures emerge over repeated updates, and
how temporal propagation influences subsequent reconstruction.

An additional motivation for adopting a QHL-based formulation, rather than
a direct local magnetic-field readout model, is that the inferred object is
not limited to a reconstructed field map alone. Because inference is
performed through an explicit Hamiltonian model, the framework can in
principle retain information about latent dynamical structure, including
shared interaction parameters such as the global coupling \(J\). In the
present study, this appears as joint inference of the spatial field and a
shared NV--NV dipolar coupling parameter within a single Bayesian
reconstruction pipeline. More broadly, such a model-based representation
may support downstream computation, control, or dynamical processing beyond
reconstruction itself, although such uses are not investigated in the
present work.

The goal of this paper is twofold. First, we establish a methodological
connection between QHL-based likelihood modeling and sequential
spatiotemporal magnetic-field-map reconstruction under an NV spin-1
Hamiltonian. Second, we demonstrate through numerical proof-of-concept
experiments that this connection is operationally meaningful for a
controlled structured synthetic field class. Specifically, we show that
local Hamiltonian-based observation models can be integrated into a
sequential reconstruction pipeline that progressively recovers maze-like
spatiotemporal structure over the analyzed frames.

The main contributions of this paper are summarized as follows:
\begin{itemize}
    \item We formulate sequential spatiotemporal magnetic-field-map
    reconstruction as a Bayesian inference problem based on local QHL
    observation models derived from a full spin-1 NV Hamiltonian.

    \item We develop a window-based reconstruction framework that combines
    Hamiltonian-induced likelihood evaluation, adaptive sequential
    updating, global field aggregation, and temporal posterior propagation.

    \item We investigate the proposed framework on controlled synthetic
    maze-like dynamic magnetic-field data and evaluate reconstruction
    accuracy under a rotating-frame NV sensing model that retains the full
    spin-1 Hilbert space, transverse strain, and dipolar interactions, with
    the dominant zero-field splitting absorbed into the reference transition
    frequency.

    \item We show that the proposed framework recovers the dominant spatial
    structure of the tested maze-like field sequence over the analyzed
    frames, while also clarifying that global coupling estimation remains
    more difficult than field reconstruction in the present full spin-1
    setting.

    \item We analyze the reconstruction process from a quantum-metrological
    viewpoint by computing classical Fisher information, quantum Fisher
    information, and leakage diagnostics for the selected controls, and we
    discuss the resulting sensitivity--leakage tradeoff relative to
    standard-quantum-limit-type scaling and idealized Heisenberg-limit
    benchmarks
    \cite{BraunsteinCaves1994,Giovannetti2006QuantumMetrology,Giovannetti2011AdvancesQuantumMetrology,Paris2009QuantumEstimation}.
\end{itemize}

The remainder of this paper is organized as follows. Section II reviews
related work. Section III introduces the problem formulation and local
NV-based observation model. Section IV presents the proposed sequential
reconstruction framework. Section V describes the experimental setup.
Section VI reports the reconstruction and metrological results.
Section VII discusses standard-quantum-limit-type scaling, idealized
Heisenberg-limit benchmarks, identifiability, limitations, and future
directions. Section VIII concludes the paper.

\section{Related Work}

QHL provides a Bayesian framework for inferring unknown physical parameters from measurement statistics generated by Hamiltonian dynamics \cite{Wiebe2014,Granade2012RobustOnlineHamiltonianLearning,Wang2017ExperimentalQHL}.
A representative formulation was established by Wiebe \textit{et al.}, who showed how Hamiltonian-induced likelihoods can be incorporated into sequential Bayesian inference even under imperfect quantum resources \cite{Wiebe2014}. This perspective is directly relevant to the present work, since our method likewise infers latent quantities through measurement probabilities generated by a structured Hamiltonian model rather than through direct observation. Closely related Bayesian methodologies have also been developed in quantum tomography and adaptive estimation. Granade \textit{et al.} studied practical Bayesian tomography from an implementation-oriented perspective \cite{Granade2016}, while Husz\'{a}r and Houlsby investigated adaptive Bayesian quantum tomography \cite{Huszar2012}. Related Bayesian and adaptive quantum-estimation ideas also appear in broader quantum parameter-estimation and tomography settings \cite{Paris2009QuantumEstimation,Helstrom1976QuantumDetection}. 
These works are relevant because they emphasize online posterior updating and adaptive measurement selection, both of which are central to our reconstruction framework. The present study differs, however, in that the unknown quantity is not a low-dimensional static parameter or a single quantum state, but a time-varying spatial magnetic field reconstructed from many local windows.
In the terminology of QHL, the exact-diagonalization backend used in the
present work plays the role of a trusted simulator. Candidate Hamiltonian
parameters are evaluated by a controlled numerical model, and the resulting
Hamiltonian-induced likelihoods are used to update the Bayesian posterior.
This is distinct from a hardware-in-the-loop implementation, but it provides a practical route to validating the reconstruction architecture before experimental deployment.
Thus, the present numerical study should be interpreted as a trusted-simulator QHL implementation: candidate Hamiltonians are evaluated by a
classical exact-diagonalization model rather than by repeated calls to a
physical quantum device. This choice is appropriate for validating the
reconstruction architecture and for isolating identifiability effects
before hardware-level deployment.

A central methodological principle in Bayesian quantum estimation is that measurement settings should be chosen according to the current uncertainty state of the estimator rather than fixed in advance. 
This principle is reflected in QHL-based adaptive inference, where candidate controls are evaluated according to their expected utility under the current posterior \cite{Wiebe2014,Huszar2012,Granade2012RobustOnlineHamiltonianLearning,Granade2016}. 
In the present work, the adaptive control mechanism follows the same information-theoretic philosophy. The measurement schedule is not treated as a static scan pattern alone; instead, local controls are selected so as to improve uncertainty reduction within each reconstruction window. Relative to prior adaptive quantum estimation settings, the main extension here is that adaptive design is embedded in a field-reconstruction pipeline rather than a single-system estimation problem.

The physical interpretation of the observation model used here is closely connected to the broader literature on quantum sensing. Degen \textit{et al.} reviewed quantum sensing based on coherent quantum dynamics and showed how external fields can be inferred through their effect on measurement statistics \cite{Degen2017}.
The metrological limits and resource scaling of such sensing protocols are closely connected to the general theory of quantum parameter estimation and quantum metrology \cite{BraunsteinCaves1994,Giovannetti2006QuantumMetrology,Giovannetti2011AdvancesQuantumMetrology,Pezze2018QuantumMetrology}.
This likelihood-based sensing viewpoint motivates our use of Hamiltonian-dependent local measurements as the primitive observation model. 
Within this broader context, nitrogen-vacancy (NV) centers in diamond provide a prominent example of field-sensitive quantum sensors. 
Representative early demonstrations include nanoscale magnetometry with single NV centers and ensembles, as well as coherent control of NV spin degrees of freedom \cite{Taylor2008DiamondMagnetometer,Maze2008NanoscaleMagneticSensing,Balasubramanian2008NanoscaleImaging,Jelezko2004SingleSpinCoherentOscillations,Childress2006CoherentDynamicsDiamond,Pham2011MagneticFieldImaging}.
Reviews by Schirhagl \textit{et al.} \cite{Schirhagl2014}, Doherty \textit{et al.} \cite{Doherty2013NVReview}, Rondin \textit{et al.} \cite{Rondin2014NVBasedMagnetometry}, Barry \textit{et al.} \cite{Barry2020SensitivityOptimization}, and Casola \textit{et al.} \cite{Casola2018} illustrate how local physical quantities can be probed through coherent quantum dynamics at small scales.
Related non-destructive coherent readout strategies have also been demonstrated in trapped-ion platforms \cite{HouSlichterLeibfried2024}, illustrating that high-fidelity quantum measurement principles extend across diverse physical implementations. The present study does not report a hardware experiment, but it does instantiate the local likelihood model with a full spin-1 NV Hamiltonian. In this sense, the present work occupies an intermediate position between abstract Hamiltonian-learning models and deployment-oriented NV sensing protocols.

Recent work has also explored learning-based strategies for improving quantum sensing protocols and adaptive quantum measurement design \cite{Kawaguchi2025DynamicRangeQCL,Granade2012RobustOnlineHamiltonianLearning}.
Kawaguchi \textit{et al.} proposed a quantum circuit learning approach that trains parameterized quantum gates to produce
a monotonic sensing response over a target field range, thereby enhancing
the dynamic range of quantum sensing in dense qubit systems
\cite{Kawaguchi2025DynamicRangeQCL}.

From the statistical viewpoint, the present work is also related to particle-based sequential Bayesian inference and nonlinear Bayesian filtering \cite{Gordon1993BootstrapFilter,LiuChen1998SMC,Doucet2001SMC,Arulampalam2002ParticleFilters,DoucetJohansen2011ParticleFiltering,Chopin2002ParticleFilterStaticModels}.
The essential idea in our method is to maintain a weighted approximation to the posterior over local field candidates and a shared global coupling parameter, updating these weights as new local measurements arrive. 
This is closely aligned with the sequential Bayesian philosophy underlying QHL and practical Bayesian tomography \cite{Wiebe2014,Granade2012RobustOnlineHamiltonianLearning,Granade2016}, while relying on the particle-filtering representation widely used in nonlinear and non-Gaussian state estimation \cite{Doucet2001SMC,Arulampalam2002ParticleFilters}.
In the present setting, this perspective is extended from parameter estimation to distributed spatiotemporal reconstruction, where posterior information must be propagated not only across repeated measurements but also across overlapping windows and successive frames.

Another relevant line of research concerns identification of quantum systems under limited observational access. 
Kato and Yamamoto studied structure identification and state initialization in spin networks with restricted access \cite{KatoYamamoto2014SpinNetwork}, providing a relevant reference point for Hamiltonian-structure inference under limited observability.
Their work is relevant because it addresses a setting in which latent interaction structure must be inferred indirectly through constrained measurements rather than through full direct access to the system. Conceptually, this is close to the present problem: our estimator does not observe the full spatiotemporal field directly, but instead infers hidden structure from local measurement windows governed by a parameterized Hamiltonian model. The connection is particularly strong for the treatment of coupling parameters. In spin-network identification, the goal is often to recover interaction structure or coupling constants from partial measurements. In our formulation, a related role is played by the global coupling parameter \(J\), which must be inferred jointly with local field variables.

A further related research direction concerns reduction of measurement cost. 
In classical signal processing, compressed sensing showed that structured signals can be reconstructed from substantially fewer measurements than dense sampling would suggest \cite{Donoho2006CompressedSensing,CandesRombergTao2006RobustUncertainty,CandesTao2006NearOptimalRecovery}.
In quantum information, related efficiency-oriented ideas appear in compressed quantum tomography and compressive quantum process tomography \cite{Gross2010,Shabani2011CompressiveQuantumProcessTomography}, as well as in shadow tomography and classical shadow methods, which estimate many properties of a quantum system from relatively few measurements \cite{Aaronson2018ShadowTomography,HuangKuengPreskill2020ClassicalShadows}.
These measurement-efficient approaches differ from the present framework in methodology. Rather than relying on randomized measurement ensembles or compressed recovery guarantees, our method emphasizes Hamiltonian-induced local likelihoods, adaptive control selection, and sequential Bayesian posterior propagation over space and time. Nevertheless, these works are relevant as neighboring approaches because they address the same high-level challenge: how to recover informative structure from limited measurements in high-dimensional estimation problems.

Taken together, the present study lies at the intersection of QHL, adaptive Bayesian inference, quantum sensing, particle-based sequential estimation, partial-access inference, and measurement-efficient reconstruction. 
From QHL, it inherits the use of Hamiltonian-induced likelihoods for parameter inference \cite{Wiebe2014,Granade2012RobustOnlineHamiltonianLearning,Wang2017ExperimentalQHL}.
From adaptive Bayesian tomography and sequential Bayesian filtering, it adopts the principle of uncertainty-aware sequential updating \cite{Granade2016,Huszar2012,Doucet2001SMC,Arulampalam2002ParticleFilters}.
From quantum sensing and NV-center magnetometry, it takes the physical interpretation that latent quantities can be inferred through their effect on coherent measurement dynamics \cite{Degen2017,Schirhagl2014,Doherty2013NVReview,Rondin2014NVBasedMagnetometry,Barry2020SensitivityOptimization,Casola2018}.
Relative to these prior directions, the main contribution of this work is to integrate these ideas into a sequential reconstruction framework for dynamic magnetic fields using an explicit full spin-1 NV Hamiltonian. In contrast to conventional QHL settings, which primarily address low-dimensional parameter estimation, the present method targets a time-evolving two-dimensional field observed only through local windows. The resulting formulation therefore extends Hamiltonian-based Bayesian inference toward sequential spatiotemporal magnetic-field reconstruction under a physically motivated NV sensing model.

\section{Problem Formulation}

We consider the problem of reconstructing a spatially distributed magnetic field that evolves over time from sequential local observations, a setting related to Bayesian inverse problems and sequential filtering \cite{KaipioSomersalo2005InverseProblems,Stuart2010BayesianInverseProblems,Doucet2001SMC}.
The target is a spatiotemporal field
\begin{equation}
B_t \in \mathbb{R}^{H \times W},
\qquad
t=0,1,\dots,T-1,
\end{equation}
defined on a two-dimensional grid of height \(H\) and width \(W\). The value \(B_t(y,x)\) represents the local magnetic-field magnitude to be inferred at spatial coordinate \((y,x)\). The reconstruction objective is to estimate the sequence
\begin{equation}
\{B_t\}_{t=0}^{T-1}
\end{equation}
from sequentially acquired local observations. Unlike static reconstruction, the present problem is not to infer each frame independently. Instead, information obtained from earlier measurements and earlier frames should be propagated forward so that the estimate becomes progressively refined over space and time.

At each frame \(t\), the field is not observed globally. Instead, measurements are acquired through local windows of fixed length \(M\). For each scan location \(s\), let
\begin{equation}
\mathcal{I}_{t,s}
=
\bigl\{
(y_{t,s,1},x_{t,s,1}),
\dots,
(y_{t,s,M},x_{t,s,M})
\bigr\}
\end{equation}
denote the ordered set of spatial sites contained in the corresponding local window. The associated local field vector is
\begin{equation}
\mathbf{b}_{t,s}
=
\bigl(
B_t(y_{t,s,1},x_{t,s,1}),
\dots,
B_t(y_{t,s,M},x_{t,s,M})
\bigr)
\in \mathbb{R}^M .
\label{eq:local_window_vector}
\end{equation}
Let
\begin{equation}
\mathcal{S}_t=\{s_1,\dots,s_{N_t}\}
\end{equation}
be the ordered set of scan locations visited at frame \(t\). Then the observation process at frame \(t\) consists of sequential local measurements on
\begin{equation}
\mathbf{b}_{t,s_1},\mathbf{b}_{t,s_2},\dots,\mathbf{b}_{t,s_{N_t}}.
\end{equation}
Because neighboring windows generally overlap, the reconstruction procedure must merge partially redundant local information into a consistent global estimate of \(B_t\).

Each local window is associated with a parameterized NV-center spin-1 measurement model, following the standard description of the NV center as a spin-1 defect with magnetic-field-sensitive coherent dynamics \cite{Doherty2013NVReview,Rondin2014NVBasedMagnetometry,Barry2020SensitivityOptimization}. 
For a local field vector
\begin{equation}
\mathbf{b}_{t,s}=(b_1,\dots,b_M)\in\mathbb{R}^M,
\end{equation}
we use an effective rotating-frame representation of the spin-1 NV
Hamiltonian. 
The dominant zero-field-splitting contribution is absorbed
into the reference transition frequency, while the local magnetic field
appears as a detuning relative to \(B_{\mathrm{ref}}\). 
The physical electron gyromagnetic ratio is negative,
\(\gamma_e<0\), with
\(|\gamma_e|=28.025\times10^9\,\mathrm{Hz/T}\).
To avoid ambiguity between the physical sign convention and the numerical
rotating-frame backend, we denote the positive magnitude used in the
simulation by
\begin{equation}
\gamma_{\mathrm{NV}}=|\gamma_e|.
\end{equation}
The real-device sensing interpretation targets the
\(|0\rangle\leftrightarrow|-1\rangle\) transition. 
The numerical backend, however, uses an equivalent rotating-frame convention in which the
Ramsey-sensitive subspace is represented as \(\{|+1\rangle,|0\rangle\}\).
This convention preserves the field-sensitivity scale set by
\(|\gamma_e|\), while retaining the full
\(\{|+1\rangle,|0\rangle,|-1\rangle\}\) manifold so that leakage can be
diagnosed explicitly.

We define the local many-body NV Hamiltonian
\begin{subequations}
\label{eq:nv_local_hamiltonian_revised}
\begin{align}
H_{\mathrm{NV}}(\mathbf{b}_{t,s},J;\Omega)
&=
\sum_{q=1}^{M} h_q(b_q;\Omega)
+
J\sum_{1\le q<r\le M}
\frac{V_{qr}}{|q-r|^{3}},
\label{eq:nv_hamiltonian_compact}
\\
h_q(b_q;\Omega)
&=
\gamma_{\mathrm{NV}} (b_q-B_{\mathrm{ref}})\Pi_q^{(+1)}
+
\gamma_{\mathrm{NV}} (b_q+B_{\mathrm{ref}})\Pi_q^{(-1)}
\nonumber\\
&\quad
+\varepsilon_q\bigl(S_{x,q}^{2}-S_{y,q}^{2}\bigr)
+\Omega H_{\mathrm{mw}}^{(q)},
\label{eq:nv_site_term}
\\
V_{qr}
&=
S_{x,q}S_{x,r}
+S_{y,q}S_{y,r}
-2S_{z,q}S_{z,r},
\label{eq:nv_dipolar_term}
\end{align}
\end{subequations}
where \(B_{\mathrm{ref}}\) is the reference field used to define the
local detuning, \(\Pi_q^{(+1)}=\lvert +1\rangle_q\langle +1\rvert\),
\(\Pi_q^{(-1)}=\lvert -1\rangle_q\langle -1\rvert\), and \(V_{qr}\)
denotes the spin--spin dipolar interaction operator.
In the present experiments, \(B_{\mathrm{ref}}=B_{\mathrm{base}}\).
The reported numerical results should therefore be interpreted under this
effective rotating-frame convention. The physical electron gyromagnetic
ratio remains negative, as required for NV-center spin physics, but the
numerical likelihood backend uses the positive sensitivity scale
\(\gamma_{\mathrm{NV}}=|\gamma_e|\). This choice keeps the stated model
consistent with the simulation backend while preserving the real-device
interpretation of NV magnetic-field sensing.

The dipolar interaction therefore follows a full \(1/r^3\) tail within
each local window rather than a nearest-neighbor truncation, consistent
with the long-range character of magnetic dipole--dipole interactions in spin systems \cite{Doherty2013NVReview}. 
In addition, the implementation includes site-dependent transverse-strain disorder
\(\varepsilon_q\), and the same full spin-1 likelihood backend is used for
posterior updating, control selection, and metrological diagnostics.

The control applied at each local measurement step is written as
\begin{equation}
u=(T,\Omega,O),
\end{equation}
where \(T\) is the evolution time, \(\Omega\) is the microwave-drive amplitude, and \(O\) is the readout observable. 
In the present numerical implementation, the observables labeled as
``Z'' and ``ZZ'' in the simulation logs are Ramsey-type effective readout
operators on the backend sensing subspace \(\{|+1\rangle,|0\rangle\}\).
We define
\begin{equation}
X_{\mathrm{eff}}^{(+1,0)}
=
|+1\rangle\langle 0| + |0\rangle\langle +1|,
\end{equation}
and use its nearest-neighbor two-site analogue
\begin{equation}
X_{\mathrm{eff}}^{(+1,0)}\otimes X_{\mathrm{eff}}^{(+1,0)}.
\end{equation}
This convention matches the numerical likelihood backend used to generate
the reported results. In a real NV sensing implementation, the same
Ramsey-readout structure would normally be associated with the
\(|0\rangle\leftrightarrow|-1\rangle\) transition; the distinction is a
rotating-frame and labeling convention rather than a change in the
Bayesian reconstruction architecture.
This choice is essential in the present sensing configuration, because a purely diagonal population-difference observable would suppress the phase sensitivity needed for magnetic-field inference.
The labels ``Z'' and ``ZZ'' are retained only to match the implementation
logs; physically, these observables correspond to Ramsey-phase readout in
the effective Ramsey sensing basis used by the numerical backend, as commonly used
in coherent spin-based NV magnetometry \cite{Taylor2008DiamondMagnetometer,Maze2008NanoscaleMagneticSensing,Degen2017,Barry2020SensitivityOptimization}.

Starting from a fixed initial state \(|\psi_0\rangle\), the ideal unitary evolution is
\begin{equation}
U(\mathbf{b}_{t,s},J;T,\Omega)
=
\exp\!\left[
-iT H_{\mathrm{NV}}(\mathbf{b}_{t,s},J;\Omega)
\right].
\label{eq:ideal_unitary}
\end{equation}
Here \(H_{\mathrm{NV}}\) is expressed in the effective frequency convention
used by the numerical backend, so that the simulated time evolution is
implemented as \(\exp[-iT H_{\mathrm{NV}}]\).
For a given observable \(O\), the induced expectation value is
\begin{equation}
\mu(\mathbf{b}_{t,s},J;u)
=
\langle\psi_0|
U(\mathbf{b}_{t,s},J;T,\Omega)^\dagger
\, O \,
U(\mathbf{b}_{t,s},J;T,\Omega)
|\psi_0\rangle ,
\label{eq:expectation_mu}
\end{equation}
and the corresponding binary-outcome probability is
\begin{equation}
p(\mathbf{b}_{t,s},J;u)
=
\frac{1+\mu(\mathbf{b}_{t,s},J;u)}{2}.
\label{eq:pplus}
\end{equation}
Strictly speaking, \(X_{\mathrm{eff}}^{(+1,0)}\) acts as a Pauli-\(X\) operator only inside the \(\{|+1\rangle,|0\rangle\}\) backend sensing subspace.
In the present numerical likelihood model, Eq.~(\ref{eq:pplus}) is used
as an effective binary Ramsey-readout probability, while population outside
the sensing subspace is not treated as a separate measurement outcome.
The resulting leakage is monitored independently through
Eq.~(\ref{eq:leakage_def}) and is not fed back into the likelihood in the
reported run.
Thus, the local magnetic-field values are not observed directly. Instead, they are inferred indirectly through the dynamical response of the NV Hamiltonian in Eq.~(\ref{eq:nv_local_hamiltonian_revised}) and the resulting measurement probability in Eq.~(\ref{eq:pplus}).

For each local window \(s\) at frame \(t\), the measurement process may be repeated over multiple adaptive steps. Let
\begin{equation}
\mathcal{D}_{t,s}
=
\bigl\{
(u_{t,s,k}, z_{t,s,k})
\bigr\}_{k=1}^{K_{t,s}}
\end{equation}
denote the measurement record for window \(s\), where \(u_{t,s,k}\) is the applied control at local step \(k\), and \(z_{t,s,k}\) is the corresponding measurement statistic. 
In the present formulation, \(z_{t,s,k}\) is modeled as a binomial count, as is standard for repeated binary-outcome measurements in quantum parameter estimation \cite{Helstrom1976QuantumDetection,Paris2009QuantumEstimation,BraunsteinCaves1994},
\begin{equation}
z_{t,s,k}
\sim
\mathrm{Binomial}
\bigl(
N_{\mathrm{shots}},
\, p(\mathbf{b}_{t,s},J;u_{t,s,k})
\bigr),
\label{eq:binomial_observation}
\end{equation}
where \(N_{\mathrm{shots}}\) is the number of repeated measurements under the same control. The corresponding single-step likelihood is
\begin{align}
\mathcal{L}_{t,s,k}
\bigl(
z_{t,s,k}\mid \mathbf{b}_{t,s},J,u_{t,s,k}
\bigr)
&=
\binom{N_{\mathrm{shots}}}{z_{t,s,k}}
p^{z_{t,s,k}}
\notag\\
&\quad\times
(1-p)^{N_{\mathrm{shots}}-z_{t,s,k}},
\label{eq:single_step_likelihood}
\end{align}
where \(p = p(\mathbf{b}_{t,s},J;u_{t,s,k})\). Assuming conditional independence across adaptive steps within the same window,
\begin{equation}
\mathcal{L}_{t,s}
(
\mathcal{D}_{t,s}\mid \mathbf{b}_{t,s},J
)
=
\prod_{k=1}^{K_{t,s}}
\mathcal{L}_{t,s,k}
\bigl(
z_{t,s,k}\mid \mathbf{b}_{t,s},J,u_{t,s,k}
\bigr).
\label{eq:window_likelihood}
\end{equation}

The complete dataset up to frame \(t\) is
\begin{equation}
\mathcal{D}_{0:t}
=
\bigcup_{\tau=0}^{t}
\bigcup_{s\in\mathcal{S}_\tau}
\mathcal{D}_{\tau,s}.
\end{equation}
Because the data are acquired in scan order, inference must be performed sequentially so that information from earlier measurements can influence later windows and later frames.

The primary inference target is the latent spatiotemporal magnetic field
\begin{equation}
\{B_t\}_{t=0}^{T-1},
\end{equation}
together with the shared global dipolar coupling parameter \(J\). Thus, the overall inference target is
\begin{equation}
\Theta
=
\bigl(
\{B_t\}_{t=0}^{T-1},\, J
\bigr).
\end{equation}
Because direct joint inference over all spatial and temporal degrees of freedom is computationally prohibitive, the problem is approached through sequential local updates. Each local dataset \(\mathcal{D}_{t,s}\) provides information primarily about the corresponding window vector \(\mathbf{b}_{t,s}\), while the shared parameter \(J\) couples inference across all windows.

At the local level, sequential Bayesian updating takes the form
\begin{equation}
p(\mathbf{b}_{t,s},J \mid \mathcal{D}_{0:t,s})
\propto
\mathcal{L}_{t,s}
(
\mathcal{D}_{t,s}\mid \mathbf{b}_{t,s},J
)
\,
p(\mathbf{b}_{t,s},J \mid \mathcal{D}_{0:t,s^-}),
\label{eq:local_bayes_update}
\end{equation}
where \(\mathcal{D}_{0:t,s^-}\) denotes the information available immediately before assimilating window \(s\) at frame \(t\). 
This local update follows the standard sequential Bayesian structure used in QHL, Bayesian tomography, and particle-filtering methods \cite{Wiebe2014,Granade2016,Doucet2001SMC,Arulampalam2002ParticleFilters}.
At the frame level, the posterior may be written abstractly as
\begin{equation}
p(B_t,J\mid \mathcal{D}_{0:t})
\propto
p(\mathcal{D}_t\mid B_t,J,\mathcal{D}_{0:t-1})
\,
p(B_t,J\mid \mathcal{D}_{0:t-1}),
\label{eq:frame_bayes_update}
\end{equation}
where \(\mathcal{D}_t=\cup_{s\in\mathcal{S}_t}\mathcal{D}_{t,s}\). In practice, this posterior is not represented exactly, but approximated through particle-based sequential inference. After processing all windows in frame \(t\), the reconstructed field is obtained by aggregating local posterior estimates:
\begin{equation}
\widehat{B}_t
=
\mathcal{A}
\bigl(
\{\widehat{\mathbf{b}}_{t,s}\}_{s\in\mathcal{S}_t}
\bigr).
\label{eq:aggregation_operator}
\end{equation}
Temporal continuity is incorporated by propagating posterior information from frame \(t-1\) to frame \(t\), following the state-space perspective common in sequential Bayesian filtering \cite{Gordon1993BootstrapFilter,Doucet2001SMC,Arulampalam2002ParticleFilters}:
\begin{equation}
p(B_t,J\mid \mathcal{D}_{0:t-1})
=
\int
p(B_t\mid B_{t-1})
\,
p(B_{t-1},J\mid \mathcal{D}_{0:t-1})
\, dB_{t-1}.
\label{eq:temporal_propagation}
\end{equation}

The present paper focuses on the methodological formulation and numerical behavior of this reconstruction problem, namely how local NV-Hamiltonian-based likelihood evaluation, global coupling inference, adaptive local updating, and frame-to-frame posterior propagation can be combined into a coherent reconstruction pipeline for dynamic two-dimensional magnetic fields.

\subsection{Metrological diagnostics}

Because the local observation model is generated by Hamiltonian evolution, the reconstruction problem can also be interpreted as a multiparameter quantum-metrology problem \cite{Helstrom1976QuantumDetection,BraunsteinCaves1994,Paris2009QuantumEstimation,Giovannetti2011AdvancesQuantumMetrology}.
For a local window, the parameter vector is
\begin{equation}
\boldsymbol{\theta}
=
(B_1,\ldots,B_M,J),
\end{equation}
where \(B_q\) denotes the local magnetic field at site \(q\), and \(J\) is the shared dipolar coupling parameter.

In the following, we refer to classical Fisher information and quantum
Fisher information as CFI and QFI, respectively.

For a binary measurement with success probability
\begin{equation}
p_+(u;\boldsymbol{\theta})
=
p(\mathbf{b},J;u),
\end{equation}
the classical Fisher information matrix associated with control \(u\) is \cite{Helstrom1976QuantumDetection,Paris2009QuantumEstimation}
\begin{equation}
F^{(\mathrm{C})}_{ab}(u)
=
N_{\mathrm{shots}}(u)
\frac{
\partial_{\theta_a}p_+(u;\boldsymbol{\theta})
\,
\partial_{\theta_b}p_+(u;\boldsymbol{\theta})
}{
p_+(u;\boldsymbol{\theta})
\left[
1-p_+(u;\boldsymbol{\theta})
\right]
},
\label{eq:cfi_matrix}
\end{equation}
where \(N_{\mathrm{shots}}(u)\) denotes the shot budget assigned to control \(u\). In the implementation, the derivatives are evaluated by finite differences around the current evaluation point.

The corresponding quantum Fisher information provides a measurement-independent local sensitivity diagnostic for the evolved state
\begin{equation}
|\psi(\boldsymbol{\theta};u)\rangle
=
U(\mathbf{b},J;T,\Omega)|\psi_0\rangle.
\end{equation}
For a pure state, the quantum Fisher information matrix is \cite{BraunsteinCaves1994,Paris2009QuantumEstimation}
\begin{align}
F^{(\mathrm{Q})}_{ab}(u)
&=
4\,\mathrm{Re}
\Big[
\langle \partial_a\psi|\partial_b\psi\rangle
-
\langle \partial_a\psi|\psi\rangle
\langle \psi|\partial_b\psi\rangle
\Big],
\label{eq:qfi_matrix}
\end{align}
where \(|\partial_a\psi\rangle=\partial_{\theta_a}|\psi(\boldsymbol{\theta};u)\rangle\). In the numerical implementation, this quantity is also estimated by finite differences.

In addition to Fisher information, we monitor leakage out of the effective sensing subspace used for Ramsey-type NV magnetometry \cite{Taylor2008DiamondMagnetometer,Degen2017,Barry2020SensitivityOptimization}.
Let
\begin{equation}
\Pi_{\mathrm{sense}}
=
\bigotimes_{q=1}^{M}
\left(
|+1\rangle_q\langle +1|
+
|0\rangle_q\langle 0|
\right)
\end{equation}
denote the projector onto the \(\{|+1\rangle,|0\rangle\}^{\otimes M}\)
backend sensing subspace.
The leakage diagnostic is defined as
\begin{equation}
L(u;\boldsymbol{\theta})
=
1-
\langle \psi(\boldsymbol{\theta};u)|
\Pi_{\mathrm{sense}}
|\psi(\boldsymbol{\theta};u)\rangle.
\label{eq:leakage_def}
\end{equation}
This quantity is not a reconstruction error by itself; rather, it measures the extent to which the full spin-1 dynamics departs from the intended effective two-level sensing manifold. 
In this sense, leakage quantifies population transfer into the unused
\(|-1\rangle\)-containing sectors of the full spin-1 Hilbert space.

The metrological diagnostics are used to characterize the selected controls and to interpret the sensitivity of the inference process. In particular, \(\sum_{q=1}^{M}F^{(\mathrm{C})}_{B_qB_q}\) summarizes local magnetic-field sensitivity, \(F^{(\mathrm{C})}_{JJ}\) summarizes coupling sensitivity, and \(L\) quantifies the cost of long-time or interaction-sensitive evolutions in terms of leakage.

\section{Proposed Sequential Reconstruction Method}

In the numerical implementation, the Bayesian posterior is represented by a particle approximation, i.e., a sequential Monte Carlo (SMC) estimator \cite{Gordon1993BootstrapFilter,LiuChen1998SMC,Doucet2001SMC,Arulampalam2002ParticleFilters}.
The term SMC therefore refers to the particle-based posterior update scheme used to approximate the local field and global coupling distributions \cite{Doucet2001SMC,DoucetJohansen2011ParticleFiltering}.
The proposed reconstruction algorithm follows the same particle-based sequential Bayesian structure as in the general formulation, but the local likelihood evaluations are now performed using the NV spin-1 Hamiltonian introduced in Section III. For each scan window \(s \in \mathcal{S}_t\) at frame \(t\), uncertainty over the local field vector \(\mathbf{b}_{t,s}\in\mathbb{R}^M\) is represented by a set of local particles
\begin{equation}
\Bigl\{
\mathbf{b}^{(i)}_{t,s}
\Bigr\}_{i=1}^{N_{\mathrm{L}}},
\end{equation}
with associated local weights
\begin{equation}
\Bigl\{
w^{(i)}_{t,s}
\Bigr\}_{i=1}^{N_{\mathrm{L}}},
\qquad
w^{(i)}_{t,s}\ge 0,
\qquad
\sum_{i=1}^{N_{\mathrm{L}}} w^{(i)}_{t,s}=1.
\end{equation}
The shared global coupling parameter \(J\) is represented by particles
\begin{equation}
\Bigl\{
J^{(j)}
\Bigr\}_{j=1}^{N_{\mathrm{G}}},
\end{equation}
with global weights
\begin{equation}
\Bigl\{
v^{(j)}_{t}
\Bigr\}_{j=1}^{N_{\mathrm{G}}},
\qquad
v^{(j)}_{t}\ge 0,
\qquad
\sum_{j=1}^{N_{\mathrm{G}}} v^{(j)}_{t}=1.
\end{equation}
To couple local and global inference, likelihoods are evaluated on candidate pairs
\begin{equation}
\bigl(
\mathbf{b}^{(i)}_{t,s}, J^{(j)}
\bigr),
\end{equation}
so that each local measurement contributes simultaneously to local field inference and to refinement of the shared coupling parameter.

At each local update step \(k\), the control \(u_{t,s,k}\) is selected adaptively from a finite candidate set \(\mathcal{U}\), following the information-gain principle used in adaptive Bayesian quantum estimation \cite{Huszar2012,Granade2016,Wiebe2014}.
Let
\begin{equation}
\rho^{(i,j)}_{t,s,k-1},
\qquad
\rho^{(i,j)}_{t,s,k-1}\ge 0,
\qquad
\sum_{i,j}\rho^{(i,j)}_{t,s,k-1}=1
\end{equation}
denote the current joint weight over candidate pairs, and let \(p^{(i,j)}(u)\) be the corresponding measurement probabilities computed from Eqs.~(\ref{eq:ideal_unitary})--(\ref{eq:pplus}). For a candidate control \(u\in\mathcal{U}\), define the mixture-averaged success probability
\begin{equation}
\bar p(u)
=
\sum_{i,j}
\rho^{(i,j)}_{t,s,k-1}
\, p^{(i,j)}(u).
\label{eq:mixture_probability}
\end{equation}
Using the entropy
\begin{equation}
\mathcal{H}(\rho)
=
-\sum_{i,j}\rho^{(i,j)}\log \rho^{(i,j)},
\end{equation}
the expected information gain of control \(u\) is
\begin{align}
\mathrm{EIG}(u)
&=
\mathcal{H}(\rho_{t,s,k-1})
-
\bar p(u)\,
\mathcal{H}\!\left(
\widetilde{\rho}^{(+)}(u)
\right)
\nonumber\\
&\quad
-
\bigl(1-\bar p(u)\bigr)
\mathcal{H}\!\left(
\widetilde{\rho}^{(-)}(u)
\right),
\label{eq:eig_def}
\end{align}
where
\begin{equation}
\begin{aligned}
\widetilde{\rho}^{(+)}_{i,j}(u)
&\propto
\rho^{(i,j)}_{t,s,k-1} p^{(i,j)}(u),
\\
\widetilde{\rho}^{(-)}_{i,j}(u)
&\propto
\rho^{(i,j)}_{t,s,k-1} \bigl(1-p^{(i,j)}(u)\bigr).
\end{aligned}
\end{equation}

For computational efficiency, the control-selection utility is evaluated
using a single-shot Bernoulli approximation to the binomial experiment.
The actual posterior update after a control is selected uses the full
binomial likelihood in Eq.~(\ref{eq:single_step_likelihood}).
A fully binomial EIG would require summing over all possible count outcomes
\(z=0,\ldots,N_{\mathrm{shots}}\), which is computationally more expensive
and is left for future refinement.
Thus, EIG measures the expected reduction in the entropy of the joint
particle posterior over local field candidates and global coupling
particles before the measurement outcome is observed, consistent with
information-theoretic adaptive design in Bayesian estimation \cite{Huszar2012,Granade2016}.

In addition to the entropy-based utility, we record metrological diagnostics for the selected controls. Let
\begin{equation}
F_B(u)
=
\sum_{q=1}^{M}
F^{(\mathrm{C})}_{B_qB_q}(u),
\qquad
F_J(u)
=
F^{(\mathrm{C})}_{JJ}(u),
\end{equation}
and let \(L(u)\) denote the leakage diagnostic in Eq.~(\ref{eq:leakage_def}). These quantities characterize the magnetic-field sensitivity, coupling sensitivity, and spin-1 leakage associated with a given control.

In the reported implementation, EIG is used as an information-theoretic
pre-screening criterion. For computational efficiency, only the top-ranked
candidate controls according to EIG are passed to a metrology-aware scoring
stage. The selected control is then
\begin{equation}
u_{t,s,k}
=
\arg\max_{u\in\mathcal{U}_{\mathrm{top}}}
\mathcal{U}_{\mathrm{score}}^{(\phi_k)}(u),
\label{eq:control_selection}
\end{equation}
where \(\mathcal{U}_{\mathrm{top}}\subset\mathcal{U}\) denotes the EIG
pre-selected candidate set, and \(\phi_k\in\{B,J\}\) denotes the phase of
the two-phase schedule.

For the \(B\)-sensitive phase, the score is
\begin{align}
\mathcal{U}_{\mathrm{score}}^{(B)}(u)
&=
\alpha_B\,\mathrm{EIG}(u)
+
\beta_B
\log\det\!\left[
F_{BB}^{(\mathrm{C})}(u)+\epsilon I
\right]
\nonumber\\
&\quad
+
\beta_{BJ}
\log\!\left[
F_{JJ}^{(\mathrm{C})}(u)+\epsilon
\right]
-
\lambda_B L(u),
\label{eq:score_B_phase}
\end{align}
where \(F_{BB}^{(\mathrm{C})}\) is the magnetic-field block of the CFI
matrix. For the \(J\)-sensitive phase, the score is
\begin{align}
\mathcal{U}_{\mathrm{score}}^{(J)}(u)
&=
\alpha_J\,\mathrm{EIG}(u)
+
\beta_{JJ}
\log\!\left[
F_{JJ}^{(\mathrm{C})}(u)+\epsilon
\right]
\nonumber\\
&\quad
+
\eta_{BJ}
\sum_{q=1}^{M}
\log\!\left[
\left|F_{B_qJ}^{(\mathrm{C})}(u)\right|+\epsilon
\right]
\nonumber\\
&\quad
-
\lambda_J L(u).
\label{eq:score_J_phase}
\end{align}
The small positive constant \(\epsilon\) is included only for numerical
regularization of nearly singular Fisher blocks.
Thus, EIG, Fisher-information diagnostics, and leakage diagnostics all
enter the implemented adaptive control policy, whereas leakage is not
included directly in the Bayesian likelihood update.

After selecting \(u_{t,s,k}\), a measurement statistic \(z_{t,s,k}\) is obtained according to the binomial model in Eq.~(\ref{eq:binomial_observation}). For each candidate pair \((i,j)\), the corresponding likelihood is
\begin{equation}
\begin{aligned}
\mathcal{L}^{(i,j)}_{t,s,k}
&=
\binom{N_{\mathrm{shots}}}{z_{t,s,k}}
\bigl(p^{(i,j)}(u_{t,s,k})\bigr)^{z_{t,s,k}}
\\
&\quad\times
\bigl(1-p^{(i,j)}(u_{t,s,k})\bigr)^{N_{\mathrm{shots}}-z_{t,s,k}}.
\end{aligned}
\end{equation}
The joint pair weights are updated by
\begin{equation}
\rho^{(i,j)}_{t,s,k}
\propto
\rho^{(i,j)}_{t,s,k-1}
\,
\mathcal{L}^{(i,j)}_{t,s,k},
\label{eq:pair_weight_update}
\end{equation}
followed by normalization. After all adaptive steps within the window, the marginal local and global weights are obtained as
\begin{equation}
w^{(i)}_{t,s}
=
\sum_{j}
\rho^{(i,j)}_{t,s,K_{t,s}},
\qquad
v^{(j)}_{t,s}
=
\sum_{i}
\rho^{(i,j)}_{t,s,K_{t,s}}.
\label{eq:marginals_from_pair}
\end{equation}
The local posterior mean is then
\begin{equation}
\widehat{\mathbf{b}}_{t,s}
=
\sum_{i=1}^{N_{\mathrm{L}}}
w^{(i)}_{t,s}\,
\mathbf{b}^{(i)}_{t,s},
\label{eq:local_posterior_mean}
\end{equation}
and the current global estimate may be summarized by
\begin{equation}
\widehat{J}_{t,s}
=
\sum_{j=1}^{N_{\mathrm{G}}}
v^{(j)}_{t,s}\,
J^{(j)}.
\label{eq:local_J_mean}
\end{equation}

After processing all windows in frame \(t\), the global field estimate \(\widehat{B}_t \in \mathbb{R}^{H\times W}\) is obtained by aggregating the window-level estimates \(\widehat{\mathbf{b}}_{t,s}\). Let \(a_{t,s}(y,x)\) denote the contribution assigned to location \((y,x)\) from window \(s\), and let \(\omega_{t,s}(y,x)\ge 0\) be the associated aggregation weight. Then
\begin{equation}
\widehat{B}_t(y,x)
=
\frac{
\sum_{s\in\mathcal{S}_t}
\omega_{t,s}(y,x)\, a_{t,s}(y,x)
}{
\sum_{s\in\mathcal{S}_t}
\omega_{t,s}(y,x)
},
\label{eq:weighted_aggregation}
\end{equation}
whenever the denominator is nonzero. In the present implementation, \(a_{t,s}(y,x)\) is obtained from the corresponding component of \(\widehat{\mathbf{b}}_{t,s}\), and \(\omega_{t,s}(y,x)\) is given by a window-dependent weighting profile. This aggregation reduces boundary artifacts and merges overlapping local estimates into a single global map.

To exploit temporal continuity, posterior information is propagated from frame \(t\) to frame \(t+1\), as in particle-filtering formulations of dynamic state estimation \cite{Gordon1993BootstrapFilter,Arulampalam2002ParticleFilters,Doucet2001SMC}.
In the particle implementation, this is realized by local perturbation of field particles together with persistence and rejuvenation of the global \(J\)-particles:
\begin{equation}
\mathbf{b}^{(i)}_{t+1,s}
=
\Pi_{\mathcal{B}}
\bigl(
\mathbf{b}^{(i)}_{t,s}
+
\boldsymbol{\eta}^{(i)}_{t,s}
\bigr),
\qquad
\boldsymbol{\eta}^{(i)}_{t,s}\sim\mathcal{N}(0,\sigma_B^2 I),
\label{eq:local_particle_propagation}
\end{equation}
and
\begin{equation}
J^{(j)}_{t+1}
=
\Pi_{[J_{\min},J_{\max}]}
\bigl(
J^{(j)}_t + \xi^{(j)}_t
\bigr),
\qquad
\xi^{(j)}_t\sim\mathcal{N}(0,\sigma_J^2).
\label{eq:J_particle_rejuvenation}
\end{equation}

\subsection{Two-phase adaptive control schedule}

To separate local field sensitivity from coupling sensitivity, the implementation uses a two-phase adaptive schedule within each local window. The first half of the adaptive steps is assigned to a \(B\)-sensitive phase, while the second half is assigned to a \(J\)-sensitive phase:
\begin{equation}
k < K_B
\quad \Rightarrow \quad
\text{\(B\)-phase},
\qquad
k \ge K_B
\quad \Rightarrow \quad
\text{\(J\)-phase}.
\end{equation}
In the experiments reported here, \(K_B=6\) and \(K_J=6\), giving \(K_{\mathrm{steps}}=12\).

The \(B\)-phase uses shorter interrogation times and primarily Ramsey-type local readout operators, because local magnetic fields enter the dominant single-site phase accumulation in NV magnetometry \cite{Taylor2008DiamondMagnetometer,Degen2017,Barry2020SensitivityOptimization}.
By contrast, the \(J\)-phase allows longer interrogation times and two-site observables, thereby increasing sensitivity to interaction-induced correlations, in analogy with Hamiltonian-parameter and coupling-structure identification problems \cite{Wiebe2014,KatoYamamoto2014SpinNetwork}.
This separation is motivated by the different dynamical roles of the local fields and the shared coupling parameter: \(B_q\) contributes directly to local detuning, whereas \(J\) affects the likelihood through weaker interaction-mediated evolution.

\begin{revtexalgorithm}[!t]{Sequential spatiotemporal magnetic-field reconstruction via NV-Hamiltonian local inference}
\label{alg:seq_qhl}
\begin{algorithmic}[1]
\REQUIRE Initial local particles \(\{\mathbf{b}^{(i)}_{0,s}\}\), global particles \(\{J^{(j)}_0\}\), and associated weights
\FOR{each frame \(t=0,1,\dots,T-1\)}
    \FOR{each scan window \(s \in \mathcal{S}_t\)}
        \STATE Construct pair weights \(\rho^{(i,j)}_{t,s,0}\)
        \FOR{each local update step \(k=1,\dots,K_{t,s}\)}
            \STATE Evaluate \(p^{(i,j)}(u)\) for candidate controls \(u\in\mathcal{U}\)
            \STATE Rank candidate controls using EIG and form \(\mathcal{U}_{\mathrm{top}}\)
            \STATE Select \(u_{t,s,k}\) using the phase-dependent score in Eq.~(\ref{eq:control_selection})
            \STATE Acquire measurement statistic \(z_{t,s,k}\)
            \STATE Update pair weights using Eq.~(\ref{eq:pair_weight_update})
        \ENDFOR
        \STATE Compute \(\widehat{\mathbf{b}}_{t,s}\) and update the global marginal over \(J\)
    \ENDFOR
    \STATE Aggregate \(\{\widehat{\mathbf{b}}_{t,s}\}_{s\in\mathcal{S}_t}\) into \(\widehat{B}_t\)
    \STATE Propagate local and global particles to frame \(t+1\)
\ENDFOR
\RETURN Reconstructed field sequence \(\{\widehat{B}_t\}_{t=0}^{T-1}\) and global coupling posterior
\end{algorithmic}
\end{revtexalgorithm}

The defining feature of the proposed method is the integration of an NV-Hamiltonian-based local likelihood model with adaptive sequential Bayesian reconstruction over space and time. This makes it suitable for studying dynamic magnetic-field estimation problems in which local measurements are generated through structured physical dynamics and reconstruction must proceed progressively from local observations.
In addition to the main H+V reconstruction setting, we also perform H-only and V-only reconstructions for comparative evaluation of scan-direction effects at the final frame.

In the numerical backend used for the reported results, the \(B\)-sensitive
phase uses a product-state Ramsey initialization, whereas the
\(J\)-sensitive phase is evaluated with a Bell-pair initialization backend.
This phase-dependent backend choice was introduced to enhance sensitivity
to the shared coupling parameter \(J\). Consequently, the \(J\)-phase
metrological diagnostics should be interpreted as diagnostics of this
coupling-enhanced backend rather than of a purely product-state protocol.

\section{Experimental Setup}

The experiments were designed to evaluate whether the proposed sequential reconstruction framework can recover a dynamic two-dimensional magnetic field from local noisy observations generated by a full spin-1 NV Hamiltonian model. The evaluation focuses on three aspects: spatial reconstruction accuracy, sequential improvement across frames, and joint inference of the shared global coupling parameter \(J\).

The ground-truth field sequence is defined on a two-dimensional grid,
\begin{equation}
B_t \in \mathbb{R}^{H \times W}, \qquad t=0,1,\dots,T-1.
\end{equation}
In the present experiments, we use \(H=W=60\) and process frames up to \(t=15\). The result reported in the main reconstruction figure corresponds to the final analyzed frame \(t=15\). 
The initial field \(B_0\) is constructed from a maze-like binary spatial pattern embedded into a continuous magnetic-field map. The base field is \(B_{\mathrm{base}}=50\,\mu\mathrm{T}\), and the field contrast is \(B_{\mathrm{amp}}=5\,\mu\mathrm{T}\), yielding a structured field in the range \(50\text{--}55\,\mu\mathrm{T}\). Temporal evolution is introduced by frame-to-frame Gaussian perturbations with standard deviation \(\sigma_{B,\mathrm{true}}=50\,\mathrm{nT}\), so that the large-scale spatial structure is preserved while mild temporal drift is added.

This field family was chosen as a controlled structured benchmark because
it contains extended spatial connectivity, sharp support boundaries, and
nontrivial long-range organization. However, it is also a highly regular
synthetic pattern. Therefore, the present experiment should be interpreted
as a proof-of-concept validation on a structured field class, not as a
comprehensive robustness study over general magnetic-field distributions.

At each frame, the field is observed through overlapping local windows of length \(M=6\). Each window defines a local field vector \(\mathbf{b}_{t,s}\in\mathbb{R}^M\), which enters the local NV Hamiltonian model. The global scan uses both horizontal and vertical passes, with \(\mathrm{stride}_{\mathrm{row}}=2\) and \(\mathrm{stride}_{\mathrm{col}}=2\). The final estimate at each frame is obtained by weighted aggregation of the local reconstructions from both scan directions.

The local likelihood model is implemented with a full spin-1 many-body
exact-diagonalization backend, reflecting the spin-1 structure of the NV
center ground-state manifold
\cite{Doherty2013NVReview,Rondin2014NVBasedMagnetometry}.
The physical electron gyromagnetic ratio is negative,
\(\gamma_e=-28.025\times10^9\,\mathrm{Hz/T}\), and its magnitude is
\(|\gamma_e|=28.025\times10^9\,\mathrm{Hz/T}\). The reference field is
taken as \(B_{\mathrm{ref}}=B_{\mathrm{base}}\). In the numerical
likelihood backend, the magnetic-field sensitivity is implemented using
the positive scale
\(\gamma_{\mathrm{NV}}=|\gamma_e|\), consistent with the rotating-frame
convention introduced in Sec.~III. Thus, the physical sign convention of
the electron spin is retained in the interpretation, whereas the numerical
backend uses the positive sensitivity scale required by the chosen
effective-frame representation.
The sensing model includes site-dependent transverse strain disorder,
enabled dipolar interactions with a full \(1/|q-r|^3\) tail, and Ramsey-type effective readout operators on the \(\{|+1\rangle,|0\rangle\}\) backend sensing subspace.
As discussed above, this backend convention is used to reproduce the reported numerical results, while the corresponding real-device interpretation targets the \(|0\rangle\leftrightarrow|-1\rangle\) transition \cite{Doherty2013NVReview,Taylor2008DiamondMagnetometer,Barry2020SensitivityOptimization}.
The \(B\)-sensitive phase uses short interrogation times
\begin{equation}
T\in\{4,8,11\}\,\mu\mathrm{s},
\end{equation}
whereas the \(J\)-sensitive phase additionally uses long interrogation times
\begin{equation}
T\in\{100,157\}\,\mu\mathrm{s}.
\end{equation}
The latter values are introduced to amplify interaction-induced phase accumulation and improve sensitivity to the shared coupling parameter, consistent with the role of interrogation time in Fisher-information-based sensing sensitivity \cite{Giovannetti2006QuantumMetrology,Degen2017,Paris2009QuantumEstimation}.

For the long-time \(J\)-sensitive controls, we use
\begin{equation}
N_{\mathrm{shots}}(100\,\mu\mathrm{s})
=
N_{\mathrm{shots}}(157\,\mu\mathrm{s})
=
300.
\end{equation}
Accordingly, the experiments should be interpreted as using a control-dependent measurement budget rather than a strictly uniform shot count across all controls.
The shot model used here is an idealized binomial model.
In practical NV devices, nuclear-assisted single-shot readout
or related non-destructive readout schemes can reduce the
effective overhead per shot by avoiding full optical
reinitialization after each measurement. Such hardware-level
readout acceleration is not modeled explicitly here, but it
would primarily affect the measurement-throughput model
rather than the Bayesian likelihood structure.

The shared global coupling parameter is fixed to \(J_{\mathrm{true}}=5.0\,\mathrm{kHz}\). For inference, the search interval is restricted to
\begin{equation}
J \in [0,10]\,\mathrm{kHz}.
\end{equation}
Each local scan window maintains \(N_{\mathrm{part}}^{\mathrm{local}}=256\) local particles, while the global coupling posterior is represented by \(N_{\mathrm{part}}^{\mathrm{global}}=256\) particles. Local temporal propagation uses \(\sigma_{B,\mathrm{dyn}}=20\,\mathrm{nT}\), local rejuvenation uses \(100\,\mathrm{nT}\) jitter, and global \(J\)-particle rejuvenation uses a jitter scale of \(1.0\,\mathrm{kHz}\).

Each local window is processed through \(K_{\mathrm{steps}}=12\) adaptive update steps. 
At each step, the algorithm first ranks candidate controls using expected information gain and then applies a phase-dependent Fisher-information and leakage-aware score to the top candidates. 
The selected control is therefore not determined by EIG alone; EIG acts as an information-theoretic pre-screening criterion, while the final control choice incorporates local magnetic-field sensitivity, coupling sensitivity, and leakage.
Metrological diagnostics are recorded for every selected control, including the diagonal magnetic-field CFI components, the coupling CFI, the corresponding QFI diagnostics, and the leakage statistics \cite{BraunsteinCaves1994,Paris2009QuantumEstimation,Giovannetti2011AdvancesQuantumMetrology}.
In the present experiments, these quantities are used primarily for post-hoc interpretation of the adaptive policy and for analyzing the sensitivity--leakage tradeoff. 
Leakage-aware likelihood correction is not applied in the reported run.
Thus, leakage is not used directly in the posterior update. 
It is, however, included in the adaptive control score to discourage controls with excessive population transfer outside the effective sensing subspace.

The reconstructed field \(\widehat{B}_t\) is compared against the ground-truth field \(B_t\) using the root-mean-square error (RMSE), Dice score, and intersection-over-union (IoU). For the global coupling parameter, we report the posterior mean and posterior standard deviation at each frame. 
The Dice score and IoU are computed after thresholding the reconstructed and true field maps into binary support patterns. They therefore quantify recovery of the maze-like spatial support rather than pointwise continuous field-amplitude accuracy. Continuous-amplitude accuracy is assessed separately using RMSE and MAE. Consequently, high Dice and IoU values should not be interpreted as evidence that the method is equally accurate for arbitrary non-binary or smoothly varying field distributions.

Table~\ref{tab:exp_params} summarizes the representative experimental configuration used throughout the present evaluation.

\begin{table}[!t]
\caption{Representative experimental parameters for the full spin-1 NV reconstruction experiment.}
\label{tab:exp_params}
\footnotesize
\setlength{\tabcolsep}{3pt}
\begin{ruledtabular}
\begin{tabular}{ll}
Parameter & Value \\
Field size \((H,W)\) & \((60,60)\) \\
Window size \(M\) & 6 \\
Number of frames \(T\) & \makecell[l]{16, indexed by\\ \(t=0,\ldots,15\)} \\
Local update steps \(K_{\mathrm{steps}}\) & 12 \\
Electron gyromagnetic ratio \(\gamma_e\) &
\makecell[l]{\(-28.025\times10^9\)\\ \(\mathrm{Hz/T}\)} \\
Magnitude \(|\gamma_e|\) &
\makecell[l]{\(28.025\times10^9\)\\ \(\mathrm{Hz/T}\)} \\
Base field \(B_{\mathrm{base}}\) & \(50\,\mu\mathrm{T}\) \\
Field amplitude \(B_{\mathrm{amp}}\) & \(5\,\mu\mathrm{T}\) \\
Reference field \(B_{\mathrm{ref}}\) & \(50\,\mu\mathrm{T}\) \\
Global coupling \(J_{\mathrm{true}}\) & \(5.0\,\mathrm{kHz}\) \\
\(J\) search range & \([0,10]\,\mathrm{kHz}\) \\
True temporal noise \(\sigma_{B,\mathrm{true}}\) & \(50\,\mathrm{nT}\) \\
Dynamic local diffusion \(\sigma_{B,\mathrm{dyn}}\) & \(20\,\mathrm{nT}\) \\
Local field jitter & \(100\,\mathrm{nT}\) \\
\(J\)-jitter & \(1.0\,\mathrm{kHz}\) \\
Local particles & 256 \\
Global \(J\)-particles & 256 \\
Scan stride \((\mathrm{row},\mathrm{col})\) & \((2,2)\) \\
Evolution times \(T\) &
\makecell[l]{\(B\)-phase: \(\{4,8,11\}\,\mu\mathrm{s}\);\\
\(J\)-phase: \(\{100,157\}\,\mu\mathrm{s}\)} \\
Drive amplitudes \(\Omega\) & \(\{5\,\mathrm{kHz}\}\) \\
Dipolar tail exponent & 3 \\
Transverse strain disorder & enabled \\
Spin--spin dipolar coupling & enabled \\
Leakage diagnostic & enabled \\
Leakage-aware control penalty & enabled in control selection \\
Leakage-aware likelihood & disabled \\
Control selection &
\makecell[l]{EIG pre-screening plus\\ FI/leakage-aware score} \\
\(J\)-phase initialization &
\makecell[l]{Bell-pair backend\\ for coupling sensitivity} \\
Nominal shots & 10000 \\
Time-dependent shots &
\makecell[l]{\(10000,3000,300\) for\\
\(4,8,11\,\mu\mathrm{s}\);\\
\(300\) for \(100,157\,\mu\mathrm{s}\)} \\
Two-phase schedule & \(K_B=6\), \(K_J=6\) \\
Metrology diagnostics & CFI, QFI, leakage \\
Likelihood evaluation & \makecell[l]{full spin-1\\ exact diagonalization} \\
Scan strategy & H+V \\
\end{tabular}
\end{ruledtabular}
\end{table}

\section{Results}

\begin{table}[t]
\caption{Frame-wise reconstruction and coupling-estimation summary for the 16-frame experiment.}
\label{tab:frame_summary}
\begin{ruledtabular}
\begin{tabular}{cccccc}
Frame \(t\) & RMSE [T] & Dice & IoU & \(\hat J\) [kHz] & \(\sigma_J\) [kHz] \\
0  & \(1.396\times10^{-6}\) & 0.9279 & 0.8654 & 4.797 & 0.068 \\
1  & \(1.337\times10^{-6}\) & 0.9379 & 0.8830 & 5.631 & 0.075 \\
2  & \(1.282\times10^{-6}\) & 0.9508 & 0.9063 & 7.499 & 0.053 \\
3  & \(1.233\times10^{-6}\) & 0.9570 & 0.9175 & 6.140 & 0.066 \\
4  & \(1.173\times10^{-6}\) & 0.9623 & 0.9273 & 6.127 & 0.075 \\
5  & \(1.118\times10^{-6}\) & 0.9701 & 0.9418 & 6.502 & 0.045 \\
6  & \(1.075\times10^{-6}\) & 0.9728 & 0.9471 & 6.399 & 0.063 \\
7  & \(1.025\times10^{-6}\) & 0.9752 & 0.9516 & 4.994 & 0.082 \\
8  & \(9.769\times10^{-7}\) & 0.9783 & 0.9576 & 6.718 & 0.074 \\
9  & \(9.367\times10^{-7}\) & 0.9824 & 0.9655 & 6.042 & 0.065 \\
10 & \(8.916\times10^{-7}\) & 0.9845 & 0.9695 & 5.420 & 0.055 \\
11 & \(8.479\times10^{-7}\) & 0.9866 & 0.9735 & 5.341 & 0.059 \\
12 & \(8.029\times10^{-7}\) & 0.9894 & 0.9790 & 5.900 & 0.104 \\
13 & \(7.708\times10^{-7}\) & 0.9902 & 0.9805 & 5.299 & 0.078 \\
14 & \(7.389\times10^{-7}\) & 0.9909 & 0.9820 & 5.725 & 0.080 \\
15 & \(7.037\times10^{-7}\) & 0.9930 & 0.9860 & 5.327 & 0.087 \\
\end{tabular}
\end{ruledtabular}
\end{table}

\begin{table}[t]
\caption{
Coupling-estimation metrological benchmark at the final checkpoint.
The finite-time values use a representative reference \(J\)-sensitive control,
\(T=157\,\mu\mathrm{s}\) with \(N=300\) shots, rather than an average over
all adaptively selected controls.
The finite-time optimal-state QFI is not computed directly; it is estimated
by multiplying the finite-time product-state QFI by the zero-time gain
\(G_{\rm opt/prod}=7.98\). The resulting value is therefore used as a
gain-extrapolated ideal-state benchmark.
}
\label{tab:J_metrology_benchmark}
\begin{ruledtabular}
\begin{tabular}{lc}
Quantity & Value \\
\(\widehat{J}_{\rm mean}\) & \(5326.9\,\mathrm{Hz}\) \\
\(J_{\rm std}^{\rm SMC}\) & \(87.0\,\mathrm{Hz}\) \\
\(|\widehat{J}_{\rm mean}-J_{\rm true}|\) & \(326.9\,\mathrm{Hz}\) \\
\(QFI_J^{\rm prod}(T\to0)\) & \(59.47\) \\
\(QFI_J^{\rm opt}(T\to0)\) & \(474.51\) \\
\(G_{\rm opt/prod}\) & \(7.98\) \\
\(\eta_M=G_{\rm opt/prod}/M\) & \(1.33\) \\
\(QFI_J^{\rm prod}(T=157\,\mu\mathrm{s})\) & \(6.199\times10^{-7}\) \\
Estimated \(QFI_J^{\rm opt}(T=157\,\mu\mathrm{s})\) & \(4.946\times10^{-6}\) \\
Product-state reference benchmark, \(N=300\) & \(73.3\,\mathrm{Hz}\) \\
Gain-extrapolated ideal-state benchmark, \(N=300\) & \(26.0\,\mathrm{Hz}\) \\
\(J_{\rm std}^{\rm SMC}/\)gain-extrapolated benchmark & \(3.35\) \\
\end{tabular}
\end{ruledtabular}
\end{table}

\subsection{Recovery of spatial magnetic-field structure}

\begin{figure}[t]
    \centering
    \includegraphics[width=\linewidth]{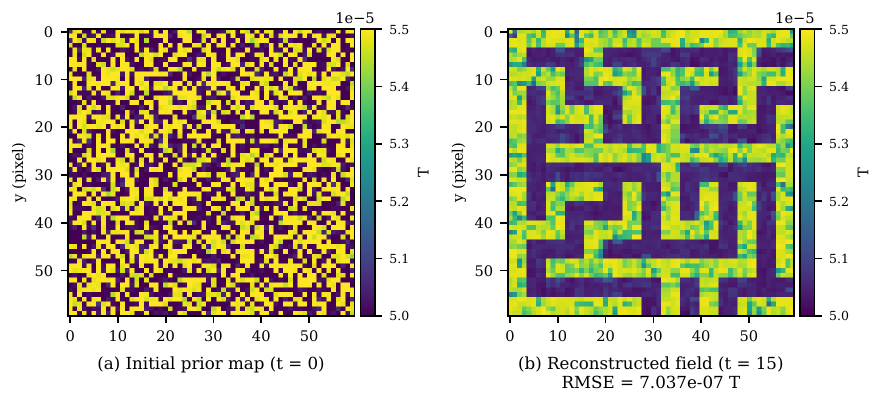}
    \caption{
    Recovery of spatial magnetic-field structure.
    (a) Initial prior map at \(t=0\), generated from a random Bernoulli
    initialization, showing no coherent global organization.
    (b) Reconstructed field at the final analyzed frame \(t=15\), obtained
    through sequential inference under the NV Hamiltonian model, with
    RMSE \(=7.037\times 10^{-7}\,\mathrm{T}\).
    The emergence of the maze-like structure supports the interpretation
    that the reconstruction is achieved through sequential inference rather
    than direct observation.
    }
    \label{fig:structure_recovery}
\end{figure}

We first examine whether the proposed method recovers coherent spatial
structure from sequential local observations. At the final analyzed frame,
the RMSE is \(7.037\times10^{-7}\,\mathrm{T}\), corresponding to
approximately \(14.1\%\) of the imposed field contrast
\(B_{\mathrm{amp}}=5\,\mu\mathrm{T}\).

The initial prior is sampled independently at each site from a Bernoulli
distribution and therefore contains no spatial correlation or large-scale
organization. In contrast, the reconstructed field in
Fig.~\ref{fig:structure_recovery}(b) exhibits a clear maze-like structure.
This indicates that the recovered pattern is not inherited from the random
prior, but emerges through local likelihood updates, overlapping-window
aggregation, scan-direction coverage, and temporal posterior propagation.

Quantitatively, the final-frame RMSE is well below the field contrast scale.
Thus, for the tested structured field sequence, the method recovers both
the dominant support structure and the continuous field magnitude at
sub-microtesla accuracy. Since the present experiment does not include a
non-Hamiltonian observation baseline with the same scan geometry, this
result is interpreted as a proof-of-concept demonstration of the complete
sequential NV-Hamiltonian reconstruction pipeline.

\subsection{Adaptive measurement behavior}

\begin{figure}[t]
    \centering
    \includegraphics[width=\linewidth]{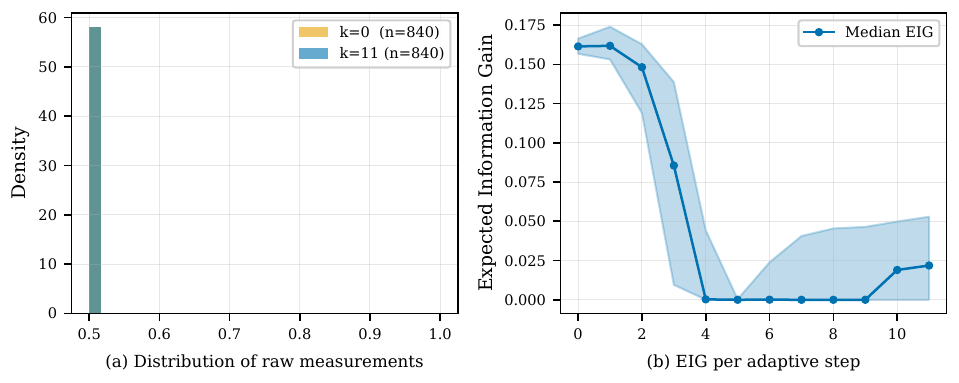}
    \caption{
    Adaptive measurement behavior.
    (a) Distribution of empirical measurement probabilities \(\hat p\) at
    the initial step (\(k=0\)) and after adaptive updates (\(k=11\)).
    (b) Expected information gain (EIG) as a function of the adaptive step
    \(k\), showing rapid decay and saturation.
    }
    \label{fig:eig_decay}
\end{figure}

Figure~\ref{fig:eig_decay} summarizes the behavior of the adaptive
measurement strategy. At \(k=0\), the empirical measurement probabilities
\(\hat p\) are broadly distributed, reflecting substantial uncertainty over
candidate local hypotheses. After adaptive updates (\(k=11\)), the
distribution becomes strongly concentrated near 0 and 1. This indicates
that the selected controls separate competing hypotheses into more
distinguishable response regimes.

The expected information gain decreases rapidly with the adaptive step and
approaches near-zero values around \(k\approx6\). Thus, most of the local
posterior reduction occurs during the early adaptive steps, whereas later
measurements primarily refine an already concentrated posterior. This
behavior is consistent with the intended Bayesian operation of the adaptive
policy: once the Hamiltonian-induced phase response is sufficiently
resolved, additional measurements provide diminishing information gain.

\subsection{Local convergence behavior}

\begin{figure}[t]
    \centering
    \includegraphics[width=\linewidth]{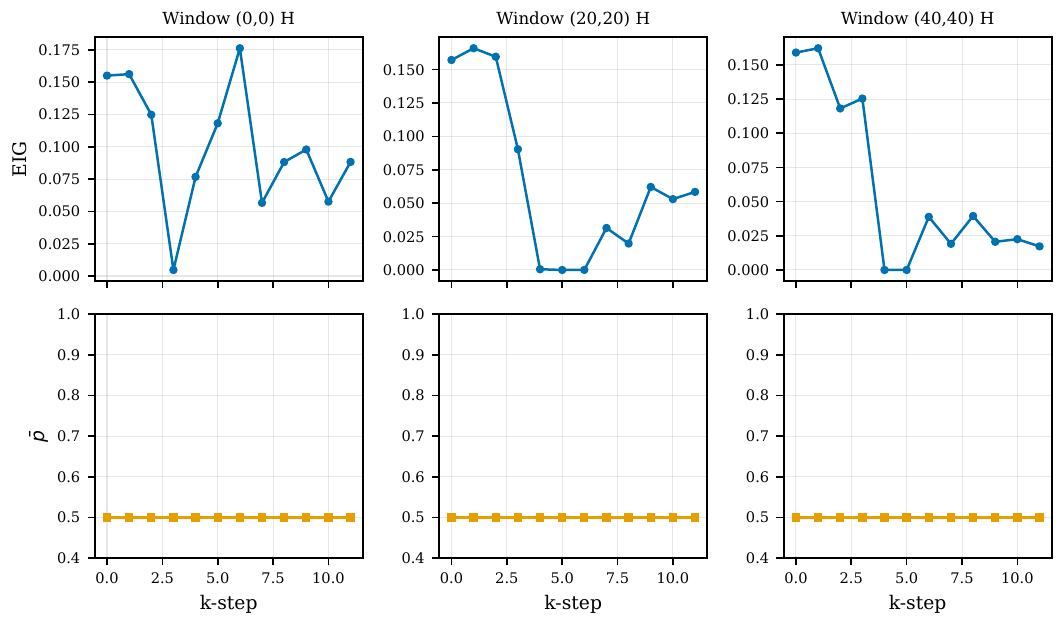}
    \caption{
    Representative \(k\)-step convergence behavior for three windows.
    Top row: expected information gain (EIG) as a function of \(k\).
    Bottom row: corresponding empirical measurement probabilities \(\bar p\).
    The EIG decreases across steps, while \(\bar p\) stabilizes after the
    early updates, consistent with local posterior concentration.
    }
    \label{fig:kstep_convergence}
\end{figure}

Figure~\ref{fig:kstep_convergence} shows representative local convergence
traces for three scan windows. All three windows exhibit decreasing EIG
across adaptive steps, but the initial magnitude and decay rate differ
across spatial locations. This variation indicates that local
identifiability depends on the field configuration within each window.

Windows containing stronger local gradients exhibit larger initial EIG and
faster convergence, whereas more homogeneous regions show lower initial
information gain and slower posterior concentration. The adaptive policy
therefore allocates effective measurement sensitivity according to local
information content, rather than applying a uniform response across the
entire field.

\subsection{Metrological diagnostics of selected controls}

We next examine the selected controls from a metrological perspective using
Fisher-information diagnostics standard in quantum parameter estimation
\cite{BraunsteinCaves1994,Paris2009QuantumEstimation,Giovannetti2011AdvancesQuantumMetrology}.
For each selected control, we record the summed magnetic-field CFI,
\begin{equation}
F_B^{(\mathrm{C})}
=
\sum_{q=1}^{M}F^{(\mathrm{C})}_{B_qB_q},
\end{equation}
the coupling CFI \(F_J^{(\mathrm{C})}=F^{(\mathrm{C})}_{JJ}\), the
corresponding QFI diagnostics, and the spin-1 leakage \(L\). These
quantities are used as diagnostics of the selected Hamiltonian evolutions,
not as direct reconstruction-error metrics.

\begin{table}[t]
\caption{Representative metrological diagnostics for selected controls in frames \(t=14\) and \(t=15\), aggregated over horizontal and vertical scan logs. \(F_B^{(\mathrm{C})}\) denotes \(\sum_{q=1}^{M}F^{(\mathrm{C})}_{B_qB_q}\).}
\label{tab:metrology_phase_summary}
\begin{ruledtabular}
\begin{tabular}{lccc}
Phase & Mean \(F_B^{(\mathrm{C})}\) & Mean \(F_J^{(\mathrm{C})}\) & Mean leakage \\
\(B\)-phase, \(k<6\) & \(5.14\times10^{13}\) & \(2.20\times10^{-7}\) & \(5.16\times10^{-8}\) \\
\(J\)-phase, \(k\ge6\) & \(2.28\times10^{15}\) & \(4.97\times10^{-6}\) & \(2.58\times10^{-6}\) \\
Enhancement & \(44.4\times\) & \(22.5\times\) & \(50.0\times\) \\
\end{tabular}
\end{ruledtabular}
\end{table}

Table~\ref{tab:metrology_phase_summary} summarizes representative
diagnostics for frames \(t=14\) and \(t=15\), separated into the
\(B\)-sensitive phase \((k<6)\) and the \(J\)-sensitive phase
\((k\ge6)\). The \(J\)-phase increases the mean summed magnetic-field CFI
from \(5.14\times10^{13}\) to \(2.28\times10^{15}\), and increases the mean
coupling CFI from \(2.20\times10^{-7}\) to \(4.97\times10^{-6}\). These
correspond to enhancement factors of approximately \(44.4\) and \(22.5\),
respectively. At the same time, the mean leakage increases from
\(5.16\times10^{-8}\) to \(2.58\times10^{-6}\), corresponding to an
approximately \(50.0\)-fold increase.

This behavior shows the sensitivity--leakage tradeoff in the full spin-1
NV model. Long-interrogation controls amplify both local-field sensitivity
and coupling sensitivity, but they also increase population leakage outside
the effective backend sensing subspace. Thus, the adaptive policy operates
in a regime where metrological sensitivity and leakage must be balanced.

We further evaluate a coupling-specific metrological benchmark at the final
checkpoint, summarized in Table~\ref{tab:J_metrology_benchmark}. For the
zero-time reference, the product-state coupling QFI is
\(QFI_J^{\rm prod}(T\to0)=59.47\), whereas the ideal optimal-state
benchmark gives \(QFI_J^{\rm opt}(T\to0)=474.51\). This corresponds to an
optimal-to-product QFI gain \(G_{\rm opt/prod}=7.98\). Normalized by the
number of spins in the window, this gives
\(\eta_M=G_{\rm opt/prod}/M=1.33\) for \(M=6\). This value is a fixed-size
ideal-state benchmark and is not a demonstration of Heisenberg scaling.

For the reference finite-time control \(T=157\,\mu\mathrm{s}\), the
product-state QFI is \(6.199\times10^{-7}\). Multiplying by the same
ideal-state gain gives an estimated optimal-state finite-time QFI of
\(4.946\times10^{-6}\). Using the single-parameter quantum
Cram\'er--Rao relation
\cite{Helstrom1976QuantumDetection,BraunsteinCaves1994,Paris2009QuantumEstimation},
\begin{equation}
\Delta J \ge \frac{1}{\sqrt{N F_J}},
\end{equation}
with \(F_J\) taken as the relevant coupling QFI, these values define a
product-state reference benchmark \(SQL_J=73.3\,\mathrm{Hz}\) and a
gain-extrapolated ideal-state benchmark of \(26.0\,\mathrm{Hz}\),
respectively. The practical SMC posterior uncertainty
\(J_{\rm std}=87.0\,\mathrm{Hz}\) is close to, but still above, the
product-state reference benchmark and remains \(3.35\times\) above the
gain-extrapolated ideal-state benchmark.

\subsection{Global coupling estimation}

\begin{figure}[t]
    \centering
    \includegraphics[width=\linewidth]{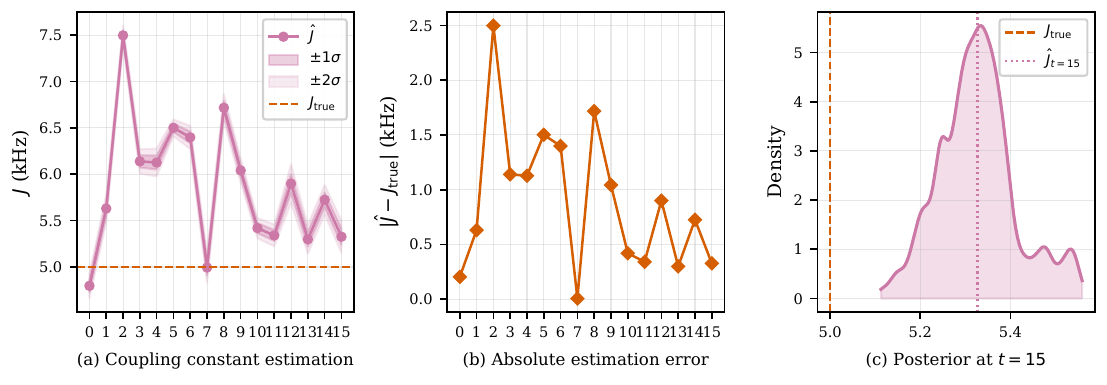}
    \caption{
    Global coupling estimation.
    (a) Posterior mean and uncertainty of \(J\) across frames.
    (b) Absolute estimation error \(|\hat J - J_{\mathrm{true}}|\).
    (c) Posterior distribution at the final frame.
    The posterior uncertainty is small, but the posterior mean remains
    frame dependent and biased relative to the true value.
    }
    \label{fig:J_tracking}
\end{figure}

Figure~\ref{fig:J_tracking} shows the global coupling posterior over the
analyzed frames. The posterior over \(J\) becomes narrow, but the posterior
mean does not converge monotonically to the true value. Instead, it remains
frame dependent, with overestimation in several intermediate frames and a
smaller positive bias at the final frame.

This behavior indicates that \(J\) is only partially identifiable under the
present sensing configuration. In particular, different combinations of
local fields and coupling values can produce similar likelihood responses,
so the coupling posterior can concentrate around a biased mode. This
distinguishes posterior concentration from unbiased parameter recovery.

At the final frame, the posterior mean is
\(\widehat{J}=5.327\,\mathrm{kHz}\) with
\(\sigma_J=0.087\,\mathrm{kHz}\), while the true value is
\(J_{\mathrm{true}}=5.0\,\mathrm{kHz}\). Equivalently, the final checkpoint
gives \(J_{\rm mean}=5326.9\,\mathrm{Hz}\) and
\(J_{\rm std}=87.0\,\mathrm{Hz}\), with an absolute bias of
\(326.9\,\mathrm{Hz}\). The ratio between the absolute bias and posterior
standard deviation is approximately
\(326.9/87.0\simeq3.76\). Hence, \(J\)-estimation is qualitatively
different from field reconstruction: the posterior width is small, but the
posterior mean remains displaced from the true coupling.

The present results do not isolate a single source of this bias. Possible
contributors include the local field--coupling likelihood geometry, the
finite control set, the particle approximation, and temporal propagation.
The important point for the present experiment is that the spatial field is
reconstructed stably while the shared interaction parameter remains more
weakly identifiable.

\subsection{Reconstruction accuracy and scan strategy}

Across the 16-frame sequence, the RMSE decreases monotonically from
\(1.396\times10^{-6}\,\mathrm{T}\) at \(t=0\) to
\(7.037\times10^{-7}\,\mathrm{T}\) at \(t=15\). The structure metrics
improve simultaneously: Dice increases from 0.9279 to 0.9930, and IoU
increases from 0.8654 to 0.9860.

\begin{figure}[t]
    \centering
    \includegraphics[width=\linewidth]{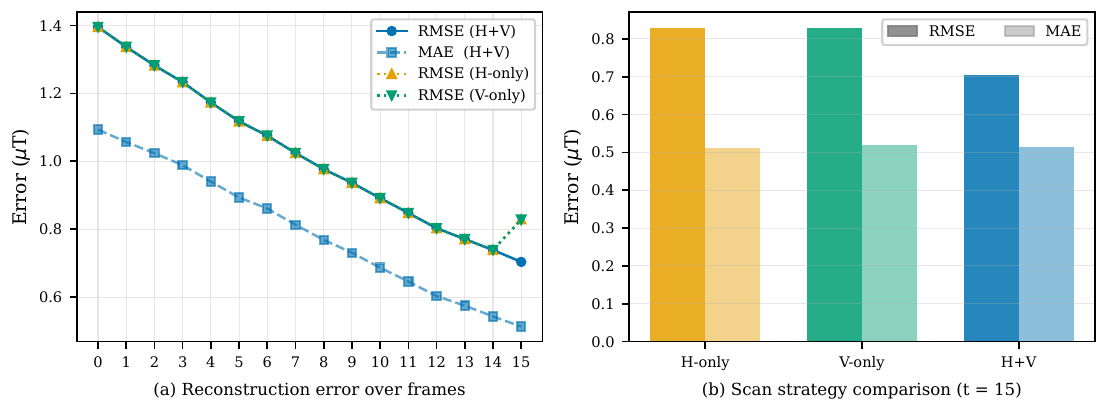}
    \caption{
    Reconstruction error and scan strategy comparison.
    (a) RMSE and MAE over frames for H+V scan.
    (b) Comparison of RMSE and MAE for H-only, V-only, and H+V scans at
    the final frame.
    }
    \label{fig:error_metrics}
\end{figure}

Figure~\ref{fig:error_metrics}(a) shows the frame-wise reconstruction error
for the H+V scan. The monotonic RMSE reduction indicates that temporal
posterior propagation improves the reconstruction over the analyzed
sequence. The improvement rate decreases at later frames, suggesting that
the dominant recoverable structure is learned early, while later frames
mainly refine the estimate. This behavior is consistent with the relatively
small temporal noise used in the simulation, which preserves the large-scale
maze-like structure across frames.

\subsection{Scan-direction comparison}

\begin{figure}[t]
    \centering
    \includegraphics[width=\linewidth]{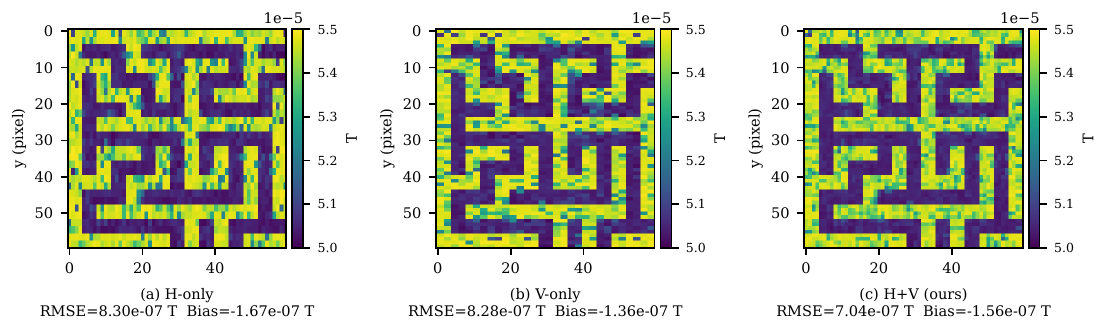}
    \caption{
    Qualitative comparison of scan strategies.
    (a) H-only reconstruction.
    (b) V-only reconstruction.
    (c) Combined H+V reconstruction.
    The H+V method yields the lowest RMSE and most faithful structural
    recovery among the tested scan strategies.
    }
    \label{fig:scan_comparison}
\end{figure}

Figure~\ref{fig:scan_comparison} compares the final-frame reconstructions
obtained using H-only, V-only, and combined H+V scans. The single-direction
scans show visible directional artifacts, reflecting reduced observability
of spatial modes poorly aligned with the scan direction. In contrast, the
H+V configuration provides complementary local windows and improves the
conditioning of the reconstruction problem.

Thus, scan-direction diversity is an important component of the proposed
pipeline. In the tested setting, combining horizontal and vertical scans
reduces anisotropic artifacts and yields the best final-frame
reconstruction accuracy.

\subsection{Scope of the comparisons}

The H-only, V-only, and H+V experiments isolate the effect of scan-direction
diversity within the proposed reconstruction pipeline. They are not intended
as a full baseline comparison against classical spatiotemporal
reconstruction methods. In particular, the present study does not include a
non-Hamiltonian observation baseline with the same scan windows, shot
allocation, temporal propagation rule, and aggregation procedure.

Accordingly, the results demonstrate the feasibility and internal behavior
of the sequential NV-Hamiltonian reconstruction pipeline, while leaving a
full ablation study to future work. Useful ablations include direct
Gaussian observation models, fixed-control variants, no-temporal-propagation
variants, and known-\(J\) versus unknown-\(J\) reconstructions.

\subsection{Summary of results}

Overall, the experiments support four main conclusions.

First, the proposed method reconstructs coherent large-scale structure in
the tested maze-like magnetic-field sequence from local NV-based
measurements. The final-frame RMSE reaches
\(7.037\times10^{-7}\,\mathrm{T}\), while Dice and IoU reach 0.9930 and
0.9860, respectively.

Second, the adaptive measurement strategy behaves consistently with
sequential Bayesian inference. EIG decreases across adaptive steps, and
representative local windows show posterior concentration after the early
updates.

Third, scan-direction diversity improves reconstruction quality. The
combined H+V scan reduces directional artifacts relative to H-only and
V-only scans and provides the best final-frame reconstruction among the
tested scan strategies.

Fourth, the shared coupling parameter \(J\) is more weakly identifiable
than the magnetic field. The final posterior uncertainty
\(J_{\rm std}=87.0\,\mathrm{Hz}\) is close to the finite-time product-state
reference benchmark \(SQL_J=73.3\,\mathrm{Hz}\), but the posterior mean
retains a bias of \(326.9\,\mathrm{Hz}\) and remains \(3.35\times\) above
the gain-extrapolated ideal-state benchmark. This indicates asymmetric
identifiability: local magnetic fields are directly imprinted through phase
accumulation, whereas the global coupling parameter influences the
likelihood indirectly through interaction-mediated dynamics.

\section{Discussion}

The results demonstrate that a local likelihood model derived from a full
spin-1 NV Hamiltonian can be embedded into a sequential reconstruction
pipeline and used to recover a coherent structured synthetic
spatiotemporal magnetic field from local noisy observations. The
significance of this result is that the recovered field structure is not
directly available from the raw local measurement statistics. Rather, the
large-scale maze-like organization emerges through repeated Bayesian
updates, overlapping-window aggregation, scan-direction coverage, and
temporal posterior propagation. This supports the central interpretation of
the proposed method as an inference-based reconstruction framework rather
than a direct local interpolation procedure.

The adaptive measurement behavior provides additional evidence that the
sequential inference mechanism is operating as intended. The expected
information gain decreases rapidly across adaptive steps, while empirical
measurement probabilities become increasingly concentrated. This indicates
that the early controls resolve most of the local posterior uncertainty,
whereas later measurements mainly refine an already concentrated posterior.
The representative window-level traces show that this convergence behavior
is not uniform across space: windows with stronger local field variation
exhibit larger initial information gain and faster posterior concentration.
Thus, the adaptive policy responds to local information content rather than
imposing a fixed measurement response across the full field.

The scan-direction comparison highlights the role of acquisition geometry
in the reconstruction problem. The H-only and V-only scans produce stronger
directional artifacts, whereas the combined H+V scan gives the lowest
final-frame RMSE and the most faithful qualitative recovery among the
tested scan strategies. This shows that reconstruction quality is governed
not only by the Hamiltonian-induced likelihood model, but also by the
spatial coverage pattern through which local observations are acquired.
In this sense, scan design plays a role analogous to measurement-setting
design in adaptive quantum estimation: it affects which spatial modes are
well constrained by the data.

From the viewpoint of Quantum Hamiltonian Learning, the present work
extends Hamiltonian-induced Bayesian likelihood modeling from isolated
parameter inference to distributed spatiotemporal reconstruction. Standard
QHL formulations typically focus on estimating a low-dimensional set of
Hamiltonian parameters for a single quantum system
\cite{Wiebe2014,Granade2012RobustOnlineHamiltonianLearning,Wang2017ExperimentalQHL}.
Here, by contrast, the Hamiltonian likelihood is evaluated repeatedly over
many local windows and is embedded into a larger reconstruction loop for a
time-evolving spatial field. The latent object is therefore not only a
small Hamiltonian parameter vector, but a dynamic field together with a
shared coupling parameter. The contribution of the present study is thus
methodological: it shows that QHL-style likelihoods can be used as local
observation models inside a sequential field-reconstruction architecture.

A key result is the asymmetry between magnetic-field reconstruction and
global coupling estimation. The local magnetic field is reconstructed
stably across the tested sequence: the RMSE decreases over frames, the
structure metrics improve, and the final field reproduces the dominant
maze-like pattern. In contrast, the shared dipolar coupling parameter \(J\)
is only partially identifiable. Its posterior becomes narrow, but the
posterior mean remains frame dependent and biased. At the final checkpoint,
the posterior has \(J_{\rm mean}=5326.9\,\mathrm{Hz}\) and
\(J_{\rm std}=87.0\,\mathrm{Hz}\), while the true value is
\(J_{\rm true}=5.0\,\mathrm{kHz}\). The absolute bias
\(326.9\,\mathrm{Hz}\) is approximately \(3.76\) times larger than the
posterior standard deviation.

This behavior separates posterior concentration from unbiased parameter
recovery. In the present likelihood geometry, different combinations of
local field values and coupling values can yield similar measurement
statistics. As a result, the particle posterior can concentrate around a
biased local mode. This does not undermine the field-reconstruction result;
rather, it identifies a specific identifiability limitation of the current
joint field--coupling inference problem. Physically, this is reasonable:
the local magnetic field directly determines the dominant phase
accumulation, whereas the coupling parameter affects the likelihood more
indirectly through interaction-mediated dynamics. More coupling-sensitive
controls, longer or richer interrogation schedules, or observables designed
to respond more directly to interaction terms will therefore be needed if
accurate recovery of \(J\) is a primary objective.

The metrological diagnostics clarify this point. The \(J\)-sensitive phase
increases both the magnetic-field CFI and the coupling CFI, but also
increases leakage out of the effective backend sensing subspace. This
sensitivity--leakage tradeoff is a characteristic feature of the full
spin-1 model used here. Longer interrogation controls improve sensitivity
to weak interaction-mediated effects, but they also enhance leakage into
unused spin sectors. The implemented adaptive policy partly accounts for
this tradeoff by including leakage-aware control scoring, while the
likelihood itself remains unchanged. Future implementations could treat
leakage more directly, for example by extending the measurement model to
include leakage-associated outcomes or by incorporating leakage-aware
likelihood corrections.

The comparison with metrological benchmarks should be interpreted as a
finite-resource diagnostic rather than as a scaling claim. For fixed
product-state probes and independent repeated measurements, the binomial
observation model gives Fisher information proportional to the number of
shots,
\begin{equation}
F^{(\mathrm{C})}\propto N_{\mathrm{shots}},
\end{equation}
and hence SQL-type uncertainty scaling,
\begin{equation}
\Delta B_{\mathrm{SQL}}
\sim
\frac{1}{\gamma_{\mathrm{NV}}T\sqrt{N_{\mathrm{shots}}}}.
\end{equation}
For \(M\) independent probes, this becomes
\begin{equation}
\Delta B_{\mathrm{SQL}}
\sim
\frac{1}{\gamma_{\mathrm{NV}}T\sqrt{M N_{\mathrm{shots}}}}.
\end{equation}
The present method should therefore not be interpreted as violating the
SQL merely because it uses Hamiltonian-based inference. Its role is instead
to improve finite-resource estimation by selecting controls with larger
posterior-dependent sensitivity.

For the shared coupling parameter, the final posterior uncertainty is close
to the finite-time product-state reference scale. Using the reference
\(J\)-sensitive control \(T=157\,\mu\mathrm{s}\) with \(N=300\) shots, the
product-state benchmark gives \(SQL_J=73.3\,\mathrm{Hz}\), while the
observed SMC posterior uncertainty is \(J_{\rm std}=87.0\,\mathrm{Hz}\).
The gain-extrapolated ideal-state benchmark gives \(26.0\,\mathrm{Hz}\),
so the observed posterior uncertainty remains \(3.35\times\) above that
benchmark. Since the finite-time optimal-state QFI is not computed
directly but obtained by applying the zero-time gain
\(G_{\rm opt/prod}=7.98\), this ideal-state value should be regarded as a
metrological reference point rather than a directly achieved performance
limit.

The Heisenberg limit provides an idealized benchmark for entangled-probe
protocols
\cite{Giovannetti2006QuantumMetrology,Giovannetti2011AdvancesQuantumMetrology,TothApellaniz2014QuantumMetrology,Pezze2018QuantumMetrology}.
In an ideal setting where the accumulated phase scales coherently with the
number of probes,
\begin{equation}
\phi_{\mathrm{HL}}\sim M\gamma_{\mathrm{NV}}BT,
\end{equation}
one obtains
\begin{equation}
\Delta B_{\mathrm{HL}}
\sim
\frac{1}{M\gamma_{\mathrm{NV}}T}.
\end{equation}
The present experiment, however, uses a fixed window size \(M=6\), a
phase-dependent backend with product-state Ramsey initialization in the
\(B\)-phase and Bell-pair initialization in the \(J\)-phase, and local or
two-site readout observables. Thus, the quantity
\(G_{\rm opt/prod}=7.98\) should be interpreted only as a fixed-\(M\)
ideal-state QFI gain for the coupling Hamiltonian. It is not a scaling
exponent and does not constitute an operational demonstration of
Heisenberg-limit scaling. Establishing SQL-to-HL scaling would require a
systematic resource study over \(N_{\mathrm{shots}}\), interrogation time,
and spin-chain length \(M\), together with explicit preparation and
measurement protocols for the relevant optimal states.

The present study is a numerical proof-of-concept evaluation, not a
hardware-level demonstration. This distinction is important but does not
weaken the methodological contribution: the goal here is to validate the
sequential inference architecture under a physically motivated full spin-1
NV likelihood model. The tested field class is a controlled synthetic
maze-like sequence with mild temporal drift, which is useful for analyzing
whether the method can recover structured spatial information under known
conditions. Broader validation should examine less regular field
morphologies, lower signal-to-noise ratios, faster temporal dynamics,
calibration drift, localization uncertainty, and model mismatch.

Several directions follow directly from these findings. First, coupling
identifiability should be improved by expanding the control set, including
richer evolution-time schedules, nonzero microwave driving, and readout
observables with stronger response to interaction terms. Second, the
effective two-level sensing manifold could be better isolated by increasing
the bias field, thereby separating the \(|+1\rangle\) and \(|-1\rangle\)
manifolds more strongly. Third, reducing the NV--NV spacing would increase
the dipolar interaction strength and may shorten the interrogation times
needed for coupling-sensitive measurements. Fourth, the computational cost
of repeated full spin-1 likelihood evaluation motivates surrogate
likelihoods, hierarchical approximations, hybrid effective/full-model
schemes, and more aggressive batching. These directions are complementary
to measurement-efficient reconstruction ideas from compressed sensing,
compressed tomography, and classical shadow methods
\cite{Donoho2006CompressedSensing,CandesTao2006NearOptimalRecovery,Gross2010,Aaronson2018ShadowTomography,HuangKuengPreskill2020ClassicalShadows}.

Overall, the present results show that Hamiltonian-based local likelihoods
derived from a physically grounded NV spin-1 model can serve as effective
building blocks for sequential reconstruction of structured synthetic
dynamic magnetic fields. The main contribution is conceptual and
methodological: a QHL-based likelihood model can be embedded into a
spatiotemporal inference architecture, enabling local quantum-dynamical
measurement models to support global field reconstruction over space and
time.

\section{Conclusion}

This paper presented a sequential spatiotemporal magnetic-field
reconstruction framework based on a nitrogen-vacancy-center spin-1
Hamiltonian likelihood model
\cite{Doherty2013NVReview,Rondin2014NVBasedMagnetometry,Barry2020SensitivityOptimization}
and adaptive Bayesian inference
\cite{Wiebe2014,Granade2016,Doucet2001SMC}. The framework combines
local window-wise Hamiltonian-induced likelihood evaluation, adaptive
sequential posterior updating, overlapping-window aggregation, temporal
posterior propagation, and joint inference of a shared global dipolar
coupling parameter. In this sense, the proposed method extends a
QHL-based likelihood formulation from low-dimensional Hamiltonian
parameter estimation toward dynamic spatial reconstruction under a
physically motivated sensing model.

Numerical proof-of-concept experiments on a controlled structured
synthetic field sequence show that the method recovers the dominant
large-scale maze-like structure from local measurement statistics that do
not directly reveal the global field. Across the 16-frame sequence, the
RMSE decreases from \(1.396\times10^{-6}\,\mathrm{T}\) at \(t=0\) to
\(7.037\times10^{-7}\,\mathrm{T}\) at \(t=15\), while the Dice score and
IoU improve to 0.9930 and 0.9860, respectively. The adaptive update
mechanism also behaves consistently with sequential Bayesian inference:
expected information gain decreases across adaptive steps, and
representative windows show stable local posterior concentration.

The scan-direction comparison further shows that combining horizontal and
vertical acquisition improves reconstruction relative to single-direction
scans in the tested setting. This indicates that spatial acquisition
geometry is an integral part of the inverse problem, alongside the
Hamiltonian likelihood model and the Bayesian update rule. The present
results therefore support the feasibility of using NV-Hamiltonian-based
local likelihoods as building blocks for structured spatiotemporal
magnetic-field reconstruction.

The global coupling parameter \(J\) exhibits a different identifiability
profile from the local magnetic field. At the final checkpoint, the
coupling posterior has
\(J_{\rm mean}=5326.9\,\mathrm{Hz}\) and
\(J_{\rm std}=87.0\,\mathrm{Hz}\), while the true value is
\(J_{\rm true}=5.0\,\mathrm{kHz}\). The absolute bias is
\(326.9\,\mathrm{Hz}\), approximately \(3.76\) times larger than the
posterior standard deviation. Thus, the posterior over \(J\) is
concentrated but biased, showing that posterior concentration alone does
not imply unbiased coupling recovery. This behavior identifies coupling
estimation as the main identifiability bottleneck of the current joint
field--coupling inference setting.

From a metrological perspective, the selected controls reveal a
Fisher-information--leakage tradeoff. Long-interrogation controls enhance
sensitivity, especially to the shared coupling parameter, while also
increasing leakage out of the effective sensing subspace. The final
coupling uncertainty \(J_{\rm std}=87.0\,\mathrm{Hz}\) is close to the
finite-time product-state reference benchmark \(SQL_J=73.3\,\mathrm{Hz}\),
but remains \(3.35\times\) above the gain-extrapolated ideal-state
benchmark. These comparisons should be interpreted as finite-resource
diagnostics, not as evidence of Heisenberg-limit scaling. Demonstrating
SQL-to-Heisenberg scaling would require systematic resource studies over
shot number, interrogation time, spin-chain length \(M\), and explicit
state-preparation and measurement protocols.

The present study is therefore best understood as a methodological
validation under controlled synthetic conditions. It demonstrates that a
full spin-1 NV Hamiltonian likelihood can be embedded into a coherent and
numerically stable sequential reconstruction architecture. Future work
should extend the validation to broader field morphologies, lower-SNR
regimes, model mismatch, and faster temporal dynamics, while improving
coupling identifiability through richer control design, more
interaction-sensitive observables, leakage-aware likelihood modeling, and
scalable likelihood-evaluation strategies. These directions provide a path
toward more realistic NV-based spatiotemporal field reconstruction and
possible hybrid classical--quantum implementations.

\begin{acknowledgments}
The authors acknowledge helpful conversations with Hideaki Yoshimura,
Mark Luo, Rei Nishimura, Naoki Yamamoto, and Rajib Shaw.
The authors thank Rajib Shaw for discussions that helped shape the
disaster-risk-reduction application perspective considered in this work.
Hiroshi Yamauchi acknowledges the support of Yoshi-aki Shimada,
Yosuke Komiyama, and Ryuji Wakikawa.
This work was partly supported by NEDO Challenge Quantum Computing
``Solve Social Issues!''.
\end{acknowledgments}

\bibliographystyle{apsrev4-2}
\bibliography{refs}

\end{document}